\documentclass[11pt,a4paper]{article}
 \usepackage{amsmath,amssymb,pifont,psfrag,bm}
  \usepackage{epsfig,graphicx}
    \graphicspath{{./Figs/}}

\begin{document}

\title{\textbf{Thresholds in FAPT: Euclid vs Minkowski}}
 \author{Alexander~P.~Bakulev$^a$\footnote{\textbf{e-mail}:bakulev@theor.jinr.ru}\\
  $^a$ \textit{\small Bogoliubov Lab. Theor. Phys., JINR} \\
       \textit{\small Jolio-Curie 6, 141980 Dubna, Russia}
}
\date{}
\maketitle

\begin{abstract}
We give a short introduction to the Analytic Perturbation Theory 
(APT)~\cite{JS95-349,JS95-357,SS96,MS96,SS98} in QCD, 
describe its problems and suggest as a tool for their resolution 
the Fractional APT (FAPT)~\cite{BMS05,BKS05,BMS06,AB08}.
We also describe shortly how to treat heavy-quark thresholds 
in FAPT.
As an applications of this technique 
we discuss 
(i) the pion form factor calculation in the  Euclidean FAPT
and 
(ii) the Higgs boson decay $H^0\to b\bar{b}$ in  Minkowskian FAPT.
We conclude with comparison of both approaches,
Euclidean and Minkowskian FAPT.
\end{abstract} 

\section{Analytic Perturbation Theory in QCD}
 \label{sec:APT}
In the standard QCD Perturbation Theory (PT) we know that 
the Renormalization Group (RG) equation $da_s[L]/dL = -a_s^2-\ldots$
for the effective coupling $\alpha_s(Q^2)=a[L]/\beta_f$ 
with $L=\ln(Q^2/\Lambda^2)$, $\beta_f=b_0(N_f)/(4\pi)=(11-2N_f/3)/(4\pi)$\footnote{%
   We use notations $f(Q^2)$ and $f[L]$ in order to specify what arguments we mean --- 
   squared momentum $Q^2$ or its logarithm $L=\ln(Q^2/\Lambda^2)$,
   that is $f[L]=f(\Lambda^2\cdot e^L)$ and $\Lambda^2$ is usually referred to $N_f=3$ region.}.
Then the one-loop solution generates Landau pole singularity,
$a_{(1)}[L] = 1/L$ with subscript $_{(l)}$ meaning $l$-loop order. 

In the Analytic Perturbation Theory (APT) we have
different effective couplings in Minkow\-skian 
(Radyushkin \cite{Rad82} and Krasnikov\&Pivovarov \cite{KP82})
and Euclidean (Shirkov\&Solovtsov \cite{SS96}) regions.
In the  Euclidean domain,
$\displaystyle-q^2=Q^2$, $\displaystyle L=\ln Q^2/\Lambda^2$,
APT generates one set of images for the effective coupling
and its $n$-th powers,  
$\displaystyle\left\{{\mathcal A}_n[L]\right\}_{n\in\mathbb{N}}$,
whereas in the Minkowskian domain,
$\displaystyle q^2=s$, $\displaystyle L_s=\ln s/\Lambda^2$,
it generates another set,
$\displaystyle\left\{{\mathfrak A}_n[L_s]\right\}_{n\in\mathbb{N}}$.
APT is based on the RG and causality 
that guaranties standard perturbative UV asymptotics 
and spectral properties.
Power series $\sum_{m}d_m a_{(1)}^m[L]$ 
transforms into non-power series 
$\sum_{m}d_m {\mathcal A}_{m}[L]$ in APT.

By the analytization in APT for an observable $f(Q^2)$
we mean the ``K\"allen--Lehman'' representation
 \begin{eqnarray}
  \label{eq:An.SD}
  \left[f(Q^2)\right]_\text{an}
   = \int_0^{\infty}\!
      \frac{\rho_f(\sigma)}
         {\sigma+Q^2-i\epsilon}\,
       d\sigma
 \end{eqnarray}
with the spectral density defined through the perturbative result,
$\rho_f(\sigma)=(1/\pi)\textbf{Im}\big[f^\text{pert}(-\sigma)\big]$.
This results in different analytic images in Euclidean and Minkowski regions
\begin{subequations}
 \label{eq:AU_n}
\begin{eqnarray}
 \label{eq:A_n}
  \mathcal A_n[L]
   &\equiv& \textbf{A}_\textbf{E}\left[a^n[L]\right]\
   =\ \int_0^{\infty}\!\frac{\rho_n(\sigma)}{\sigma+Q^2}\,d\sigma\,,~\\
 \label{eq:U_n}  
  \mathfrak A_n[L_s] 
   &\equiv& \textbf{A}_\textbf{M}\left[a^n[L]\right]\
   =\ \int_s^{\infty}\!\frac{\rho_n(\sigma)}{\sigma}\,d\sigma\,.~
 \end{eqnarray}
\end{subequations}
Then in the one-loop approximation spectral density is
$\rho_1^{(1)}(\sigma)=1/\left[\ln^2(\sigma/\Lambda^2)+\pi^2\right]$
and analytic couplings are
\begin{eqnarray}
 \label{eq:AU_1L}
  \mathcal A_1^{(1)}[L]
   = \frac{1}{L} - \frac{1}{e^L-1}\,,~~~~~
  \mathfrak A_1^{(1)}[L_s] 
   = \frac{1}{\pi}\,\arccos\frac{L_s}{\sqrt{\pi^2+L_s^2}}\,,~
 \end{eqnarray}
whereas spectral density for the analytic images of the higher powers 
($n\geq2, n\in\mathbb{N}$) is 
$\rho_n^{(1)}(s)=\left(1/(n-1)!\right)\left(-d/d L_s\right)^{n-1}\rho_1^{(1)}(s)$
and analytic images of the higher powers are:
\begin{eqnarray}
 \label{eq:AU.rec}
 {\mathcal A^{(1)}_n[L] \choose \mathfrak A^{(1)}_n[L_s]}
  &=& \frac{1}{(n-1)!}\left( -\frac{d}{d L}\right)^{n-1}
      {\mathcal A_{1}[L] \choose \mathfrak A_{1}^{(1)}[L_s]}\,.
\end{eqnarray}
    
\section{Fractional APT}
 \label{sec:FAPT}
\begin{figure}[t]
 \centerline{\includegraphics[width=0.45\textwidth]{
             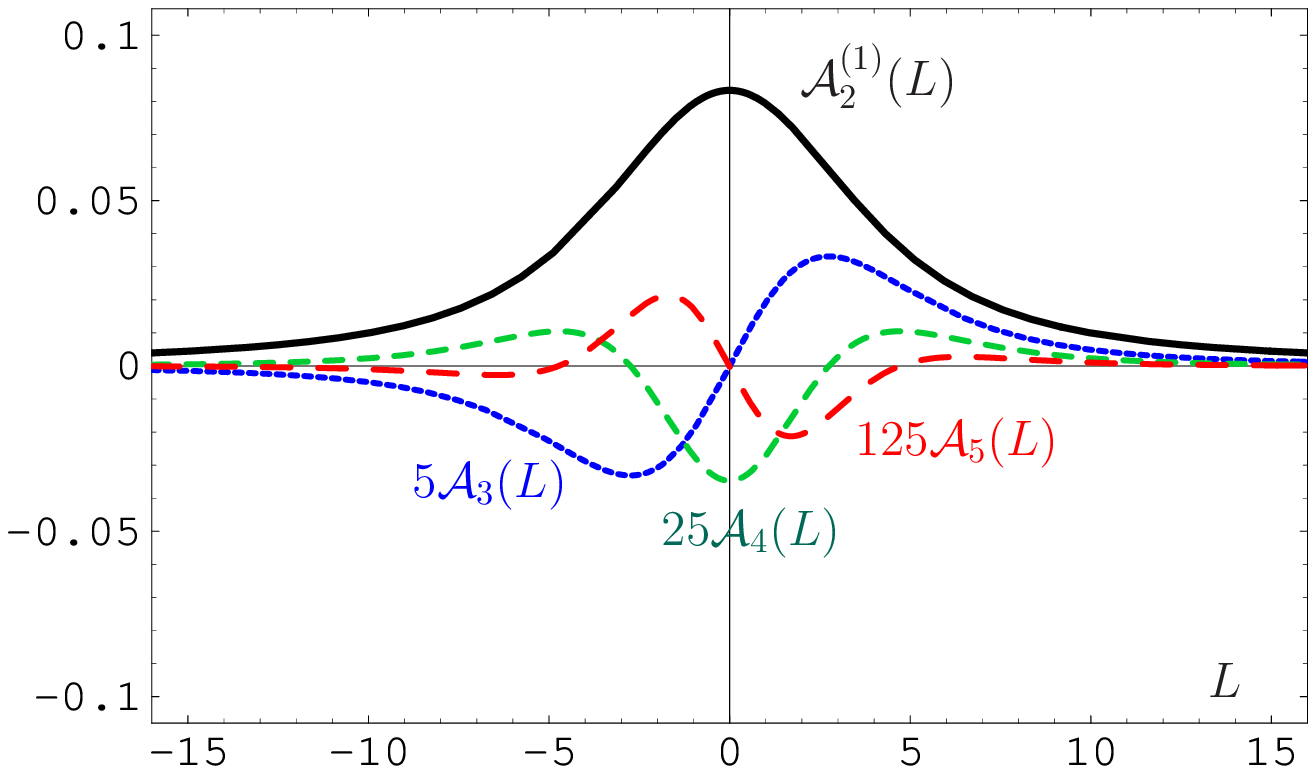}~~~%
             \includegraphics[width=0.45\textwidth]{
             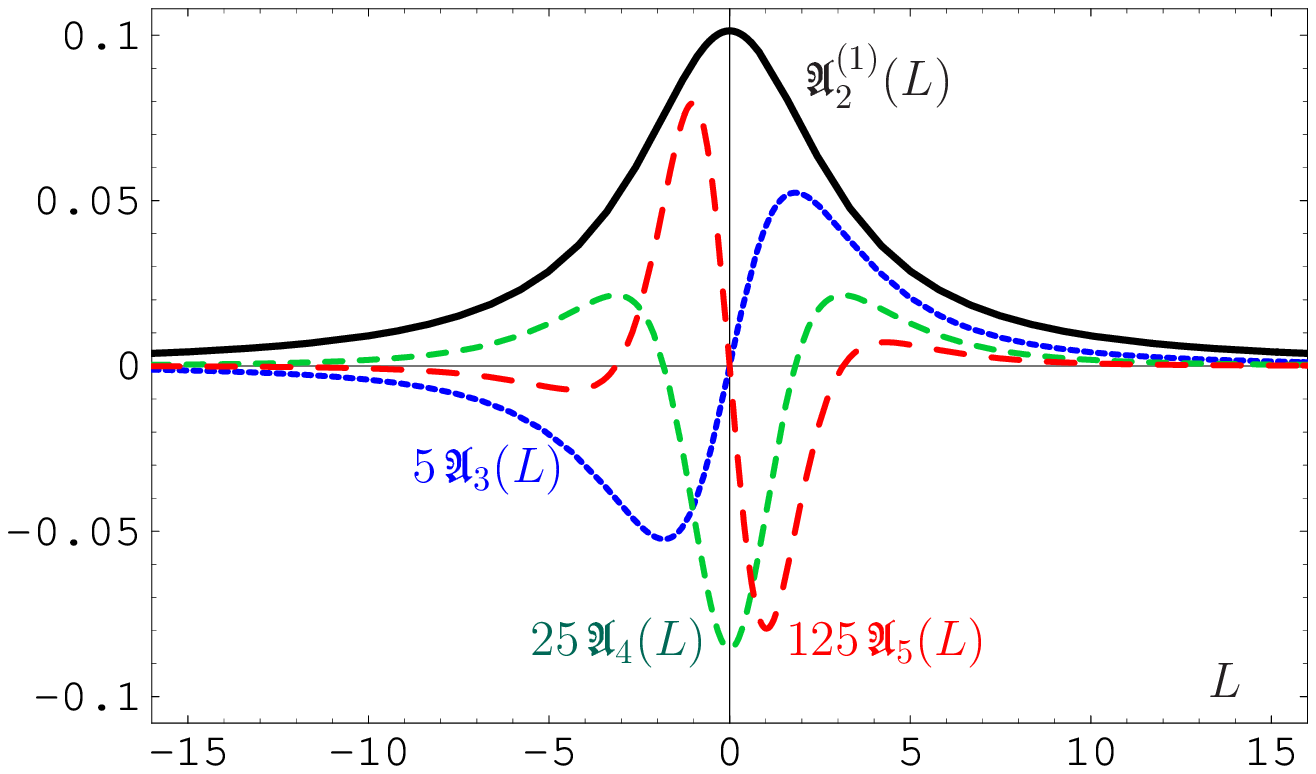}}
  \caption{\small Graphics of ${\mathcal A}_{\nu}[L]$ (left panel)
  and ${\mathfrak A}_{\nu}[L]$ (right panel)
  for integer $\nu=2, 3, 4, 5$. In order to show all curves 
  on the same panel we scale different curves by factors $5^{\nu-2}$. 
  \label{fig:UA_2345}}
\end{figure}

In the standard QCD PT we have also:
\begin{itemize}
 \item the factorization procedure in QCD  
    that gives rise to the appearance of logarithmic factors of the type: 
     $a^\nu[L]\,L$;~\footnote{%
     First indication that a special ``analytization'' procedure
is needed to handle these logarithmic terms appeared in~\cite{KS01},
where it has been suggested that one should demand 
the analyticity of the partonic amplitude as a \textit{whole}.}
 \item the RG evolution 
     that generates evolution factors of the type: 
     $B(Q^2)=\left[Z(Q^2)/Z(\mu^2)\right]$ $B(\mu^2)$, 
     which reduce in the one-loop approximation to
     $Z(Q^2) \sim a^\nu[L]$ with $\nu=\gamma_0/(2b_0)$ 
     being a fractional number.
\end{itemize}
That means that in order to analytize perturbative QCD expressions 
we need to construct analytic images of new functions:
$\displaystyle a^\nu,~a^\nu\,L^m, \ldots$\,.
The result of this procedure is just 
the Fractional APT (FAPT).
Formally speaking, in order to construct FAPT
we need to use formulas (\ref{eq:AU_n})
with substituting spectral densities with integer indices $n$, 
namely $\rho_{n}(\sigma)$,
by corresponding spectral densities $\rho_{\nu}(\sigma)$
with fractional indices $\nu$.
But in the one-loop approximation FAPT can more effectively 
be constructed
using recursive relations (\ref{eq:AU.rec}), 
see for details in~\cite{BMS05,BMS06}.
The explicit expressions for couplings in Euclidean,
$\mathcal A_{\nu}^{(1)}[L]$,
and Minkowskian, $\mathfrak A_{\nu}^{(1)}[L]$,
domains are
\begin{eqnarray}
 {\mathcal A}_{\nu}^{(1)}[L] 
  = \frac{1}{L^\nu} 
  - \frac{F(e^{-L},1-\nu)}{\Gamma(\nu)}\,;\quad
 {\mathfrak A}_{\nu}^{(1)}[L] 
  = \frac{\text{sin}\left[(\nu -1)\arccos\left(L/\sqrt{\pi^2+L^2}\right)\right]}
         {\pi(\nu -1) \left(\pi^2+L^2\right)^{(\nu-1)/2}}\,.~
\end{eqnarray}
Here $F(z,\nu)$ is reduced Lerch transcendental function,
which is an analytic function in $\nu$.
\begin{figure}[h]
 \centerline{\includegraphics[width=0.45\textwidth]{
             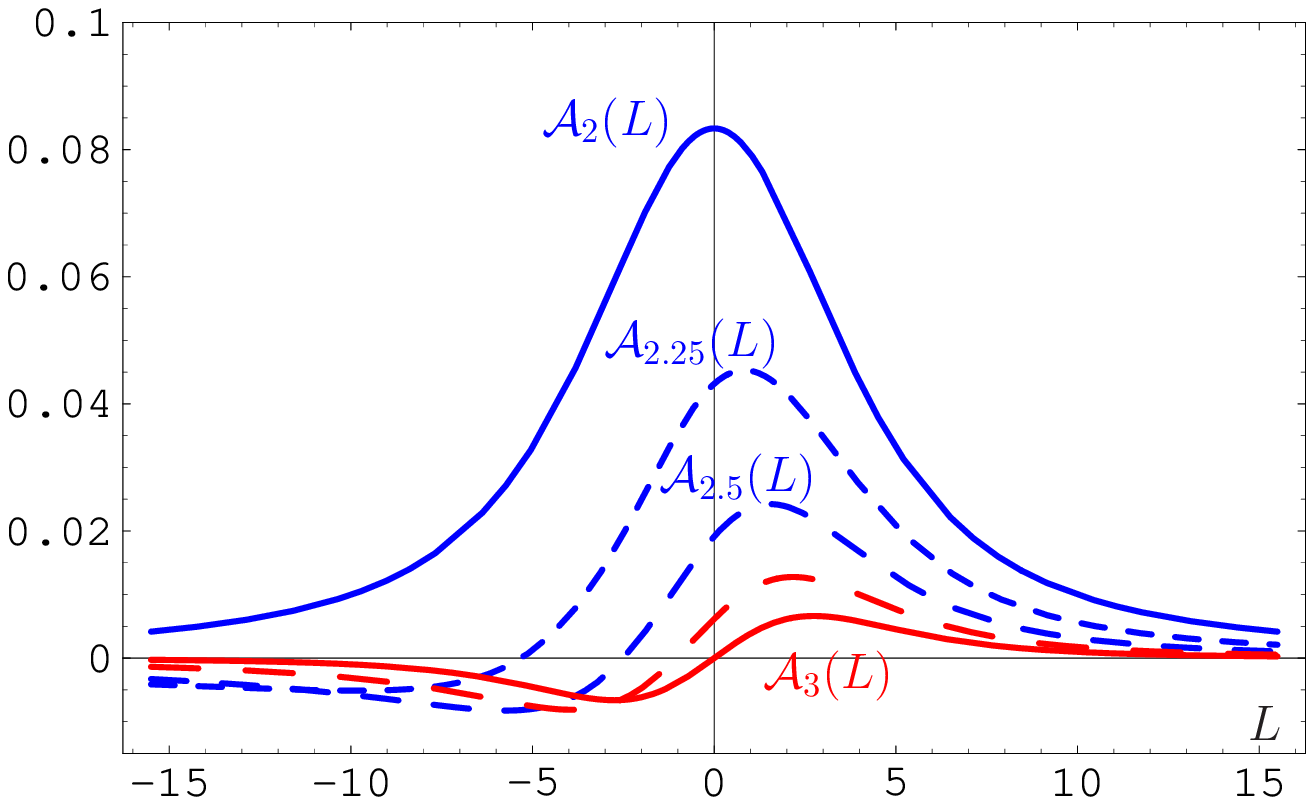}~~~%
             \includegraphics[width=0.45\textwidth]{
             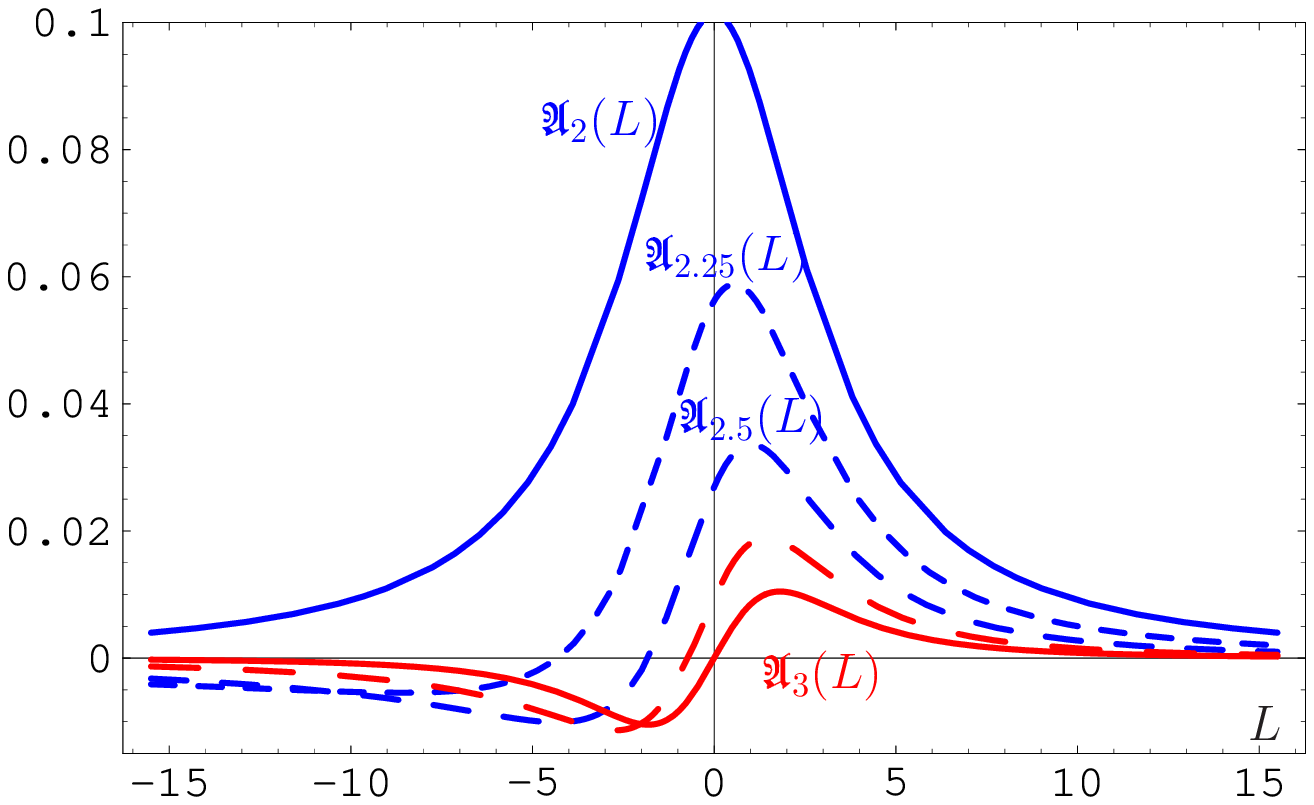}}
  \caption{\small Graphics of ${\mathcal A}_{\nu}^{(1)}[L]$ (left panel)
  and ${\mathfrak A}_{\nu}^{(1)}[L]$ (right panel)
  for fractional $\nu\in\left[2,3\right]$.
  \label{fig:U23_A23}}
\end{figure} 
Interesting to note that ${\mathcal A}_\nu^{(1)}[L]$ appears to be 
an entire function in $\nu$, 
whereas ${\mathfrak A}_{\nu}^{(1)}[L]$ 
is determined completely in terms of elementary functions.
These couplings have the following properties:
\begin{subequations}
 \label{eq:A.U}
\begin{eqnarray}
 \label{eq:AU.Prop.1}
  \mathcal A_{0}^{(1)}[L] 
   \!\!&\!\!=\!\!&\!\!\mathfrak A_{0}^{(1)}[L]=1\,;\\
 \label{eq:AU.Prop.2}
  \mathcal A_{-m}^{(1)}[L]
   \!\!&\!\!=\!\!&\!\!L^m
   ~\text{for}~m\in\mathbb{N}\,;~
   \mathfrak A_{-1}^{(1)}[L]=L\,,~
   \mathfrak A_{-2}^{(1)}[L]=L^2-\frac{\pi^2}{3}\,,~
   \mathfrak A_{-3}^{(1)}[L]=L^3-\pi^2L\,,\,\ldots\,;~~~~~\\
 \label{eq:AU.Prop.3}
 {\mathcal A_{m}^{(1)}[L]
  \choose
  \mathfrak A_{m}^{(1)}[L]}
   \!\!&\!\!=\!\!&\!\! (-1)^{m}
    {\mathcal A_{m}^{(1)}[-L]
     \choose
     \mathfrak A_{m}^{(1)}[-L]} 
     ~\text{for}~m\geq2\,, m\in\mathbb{N}\,;
\end{eqnarray}
\begin{eqnarray} \label{eq:AU.Prop.4}
  \mathcal A_{m}^{(1)}[\pm\infty]
   \!\!&\!\!=\!\!&\!\!\mathfrak A_{m}^{(1)}[\pm\infty]=0
    ~\text{for}~m\geq2\,,\ m\in\mathbb{N}\,;\\
 \label{eq:AU.Prop.5}
  \mathcal D^{k}
   {\mathcal A_{\nu}^{(l)}
    \choose
    \mathfrak A_{\nu}^{(l)}} 
    \!\!&\!\!=\!\!&\!\! 
  \frac{d^k}{d \nu^k}
   {\mathcal A_{\nu}^{(l)}
    \choose
    \mathfrak A_{\nu}^{(l)}}
    =
   {\textbf{A}_\textbf{E}
    \choose
    \textbf{A}_\textbf{M}}\,\Big[a_{(l)}^\nu\ln^k(a_{(l)})\Big] 
    = 
   {\textbf{A}_\textbf{E}
    \choose
    \textbf{A}_\textbf{M}}\,\left[\frac{d^k}{d \nu^k}a_{(l)}^\nu\right]\,.~~~
\end{eqnarray}
\end{subequations}
We display graphics of $\mathcal A_{\nu}^{(1)}[L]$ and $\mathfrak A_{\nu}^{(1)}[L]$
in Fig.\ \ref{fig:U23_A23}:
one can see here a kind of distorting mirror on both panels.
Next, in Fig.\ \ref{fig:UA_2345} we show graphics for $\nu=2, 3, 4, 5$.
Here we can trace the partial values
\begin{subequations}
\begin{eqnarray}
 {\mathcal A }_{2}^{(1)}[0] = \frac{1}{12}\,,~~
 {\mathcal A }_{4}^{(1)}[0] = \frac{-1}{720}\,,~~
 {\mathcal A }_{3}^{(1)}[0] = {\mathcal A }_{5}^{(1)}[0]=0\,;~~\\
 {\mathfrak A }_{2}^{(1)}[0] = \frac{1}{\pi^2}\,,~~
 {\mathfrak A }_{4}^{(1)}[0] =-\frac{1}{3\pi^4}\,,~~
 {\mathfrak A }_{3}^{(1)}[0] = {\mathfrak A }_{5}^{(1)}[0] =0\,.
\end{eqnarray}
\end{subequations}
\\
Graphics for ${\mathcal A}_{\nu}[L]$ as functions of $\nu$ at fixed values of $L$
can be found in our last papers~\cite{BMS05}.
We compare the basic ingredients of FAPT in 
Table \ref{tab:PT.APT.FAPT} with their counterparts 
in conventional PT and APT.
\begin{table}[htb]\hfil
 \begin{tabular}{ccccc}\hline\hline
  Theory  & PT        & APT    & FAPT(E)   & FAPT $\vphantom{^{\int}}$
  \\ \hline
  Space   & $\Big\{a^\nu\Big\}_{\nu\in\mathbb{R}}\vphantom{^{\big|}_{\big|}}$
                      & $\Big\{{\mathcal A}_m\Big\}_{m\in\mathbb{N}}$
                               & $\Big\{{\mathcal A}_\nu\Big\}_{\nu\in\mathbb{R}}$
                                        & $\Big\{{\mathfrak A}_\nu\Big\}_{\nu\in\mathbb{R}}$
  \\ \hline
  Series expansion
          &~$\sum\limits_{m}f_m\,a^m[L]\vphantom{^{\big|}_{\big|}}$~
                      &~$\sum\limits_{m}f_m\,{\mathcal A}_m[L]$~
                               &~$\sum\limits_{m}f_m\,{\mathcal A}_m[L]$~
                                        &~$\sum\limits_{m}f_m\,{\mathfrak A}_m[L]$~
  \\ \hline
  Inverse powers
          & $\left(a[L]\right)^{-m}\vphantom{^{\big|}_{\big|}}$
                      & ~\text{---}~
                               & ${\mathcal A}_{-m}[L]=L^m$
                                        & ${\mathfrak A}_{-m}[L]=L^m+O(\pi^2)$
  \\ \hline
  Index derivative
          & $a^{\nu}\ln^{{k}}{a}\vphantom{^{\big|}_{\big|}}   $
                      & ~\text{---}~
                               & $\mathcal D^{k}\mathcal A_\nu$
                                        & $\mathcal D^{k}\,\mathfrak A_{\nu}$
  \\ \hline
  Logarithms
          & $a^{\nu}L^{k}\vphantom{^{\big|}_{\big|}}$
                      & ~\text{---}~
                              & $\mathcal A_{\nu-k}$
                                        & $\mathfrak A_{\nu-k}$
  \\ \hline\hline\vspace*{-7mm}
  \end{tabular}\hfil
\caption{\small Comparison of PT, APT, and FAPT in Euclidean 
 (FAPT(E), $\displaystyle L=\ln\left(Q^2/\Lambda^2\right)$),
 and Minkowski (FAPT(M), $\displaystyle L=\ln\left(s/\Lambda^2\right)$)
 domains in the one-loop approximation.
 In the row, named `Inverse powers', 
 we put ${\mathfrak A}_{-m}[L]=L^m+O(\pi^2)$
 that encodes just Eq.\ (\ref{eq:AU.Prop.2}). 
 \label{tab:PT.APT.FAPT}}  
\end{table}

\section{Developments of FAPT}
 \subsection{Two-loop coupling}
  \label{subsec:2Loop.FAPT}
In this section we want to show how good is FAPT
in approximating the two-loop analytic coupling expressions
by expanding them in non-power series in terms 
of one-loop analytic couplings.
To this end, we remind 
the two-loop equation for the normalized coupling $a_{(2)}=b_0\,\alpha/(4\pi)$: 
\begin{eqnarray}
 \frac{d a_{(2)}[L]}{dL}
    = - a_{(2)}^2[L]\left[1 + c_1\,a_{(2)}[L]\right]
      \quad \text{with}~c_1\equiv\frac{b_1}{b_0^2}\,.
\end{eqnarray}
\begin{figure}[b]
 \begin{minipage}{\textwidth}
  \centerline{\includegraphics[width=0.45\textwidth]{
              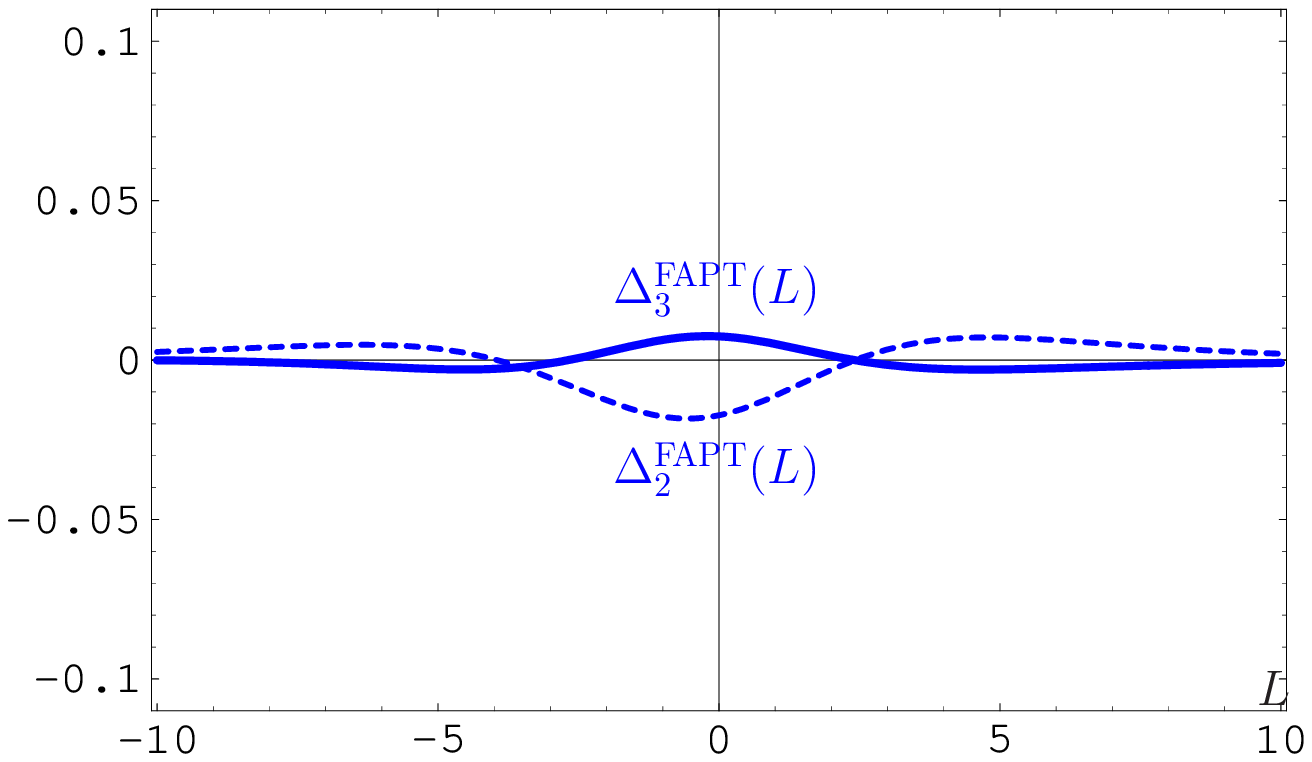}~~~%
              \includegraphics[width=0.445\textwidth]{
              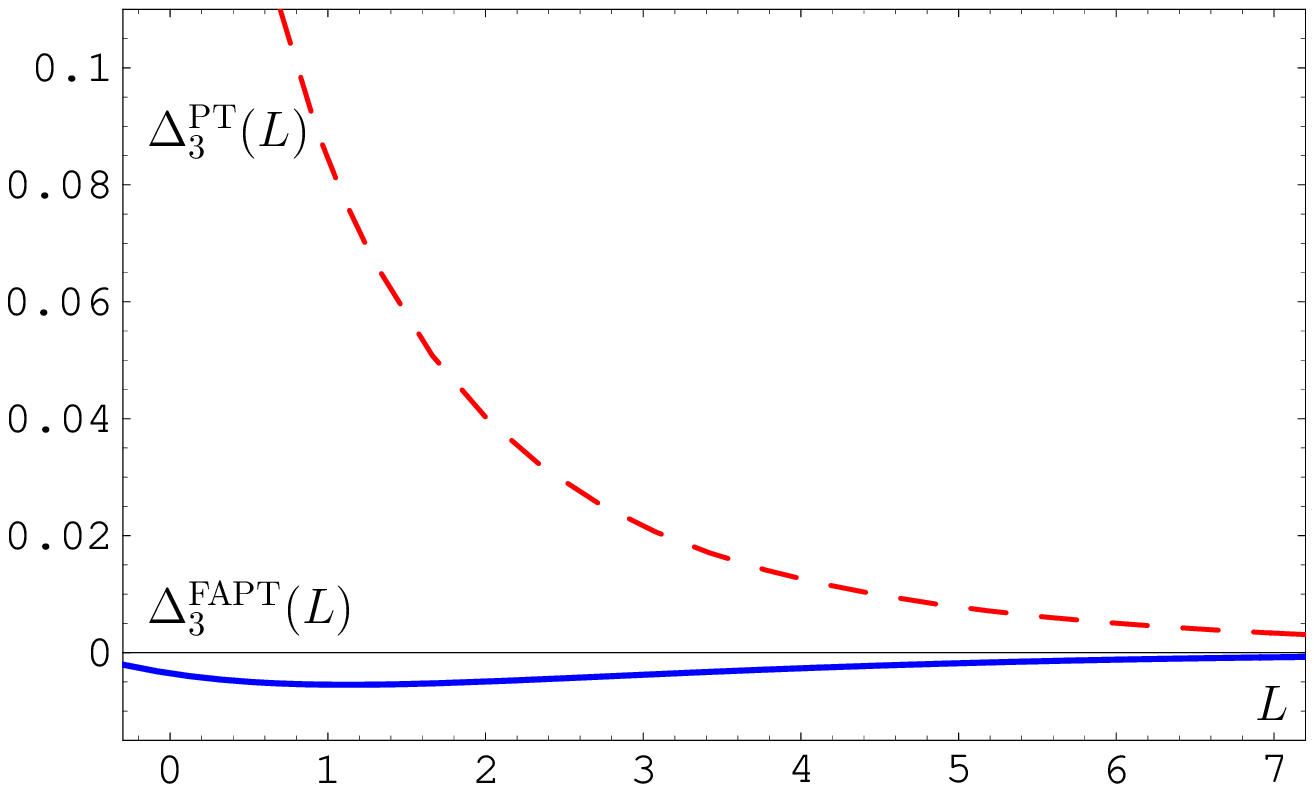}}%
  \caption{\small Left panel: Comparison of relative errors 
   $\Delta_2^\text{FAPT}[L]$ (dotted line) and 
   $\Delta_3^\text{FAPT}[L]$ (solid line) in FAPT.
   Right panel: Comparison of relative errors 
   $\Delta_3^\text{PT}[L]$ (dashed line) in standard PT 
   and $\Delta_3^\text{FAPT}[L]$ (solid line) in FAPT.
  \label{fig:FAPT23_PT3}}
\end{minipage}
\end{figure}
\\
The exact solution $a_{(2)}[L]$ of this equation satisfies the following functional equation:
\begin{eqnarray}
 \frac{1}{a_{(2)}[L]} + c_1 \ln\left[\frac{a_{(2)}[L]}{1+c_1 a_{(2)}[L]}\right] 
  = L\,.
\end{eqnarray}
Its exact solution is known~\cite{Mag99} to be 
\begin{eqnarray}
 \label{eq:App-Exactsolution.2L}
 a_{(2)}[L] =
 -\frac{1}{c_1} \frac{1}{1+W_{-1}(z_W[L])}\, ,
 \end{eqnarray}
where
$z_W[L]=\left(1/c_1\right) \exp\left(-1+i\pi-L/c_1\right)$
and $W_{k}$, $k=0,\pm 1,\ldots $, 
denote different branches of the Lambert function 
$W(z)$, 
defined through functional equation
$z=W(z)\,e^{W(z)}$.
Review of its properties can be found in ~\cite{CGHJK96,Mag99,Mag00}.

To construct FAPT for the two-loop quantities
one should use formulas (\ref{eq:AU_n})
with substituting the spectral densities $\rho_{n}(\sigma)$
by the corresponding two-loop spectral densities $\rho_{\nu}^{(2)}(\sigma)$,
defined by
\begin{eqnarray}
 \label{eq:Spec.Den.2L.nu}
  \rho_{\nu}^{(2)}(\sigma) &=&\frac{1}{\pi}
   \textbf{Im}{}
    \left[a_{(2)}^{\nu}[L-i\pi] 
    \right]\,.
\end{eqnarray}
Then analytic images 
in Euclidean,
$\mathcal A_{\nu}^{(2)}[L]$,
and Minkowskian, $\mathfrak A_{\nu}^{(2)}[L]$,
domains 
are defined through  
\begin{eqnarray}
 \label{eq:AU_n.2L}
  \mathcal A_\nu^{(2)}[L]
   = \int_0^{\infty}\!\frac{\rho_\nu^{(2)}(\sigma)}{\sigma+Q^2}\,d\sigma\,,~~~
  \mathfrak A_\nu^{(2)}[L_s] 
   = \int_s^{\infty}\!\frac{\rho_\nu^{(2)}(\sigma)}{\sigma}\,d\sigma\,.
 \end{eqnarray}

We can also expand $a_{(2)}[L]$ 
in terms of $a_{(1)}[L]=1/L$ 
with inclusion of terms $O(a_{(1)}^{3})$:
\begin{eqnarray}
 a_{(2)}[L]
   = a_{(1)}[L]
   + c_1\,a_{(1)}^2[L]\,\ln\,a_{(1)}[L]
   + c_1^2\,a_{(1)}^3[L]
        \left(\ln^2 a_{(1)}[L]
            + \ln\,a_{(1)}[L]
            -1 \right) 
       + \ldots
\end{eqnarray}
and then produce 
analytic version of this expansion
\begin{eqnarray}
 {\cal A }_{1}^{(2);\text{FAPT}}[L]
  &=& {\cal A }_{1}^{(1)}
      + c_1\,{\cal D}\,{\cal A }_{\nu=2}^{(1)}
      + c_1^2\left({\cal D}^{2}+{\cal D}-1\right)
        {\cal A }_{\nu=3}^{(1)}
      + \ldots\,.
\end{eqnarray} 
In Fig.\ \ref{fig:FAPT23_PT3} 
we demonstrate nice convergence of this expansion 
using relative errors 
of the 2- and 3-term approximations:
\begin{eqnarray}
  \Delta_2^\text{FAPT}[L] 
    \!&\!=\!&\! 1 
     - \frac{{\cal A}_{1}^{(1)}[L] + c_1\,{\cal D}{\cal A}_{\nu=2}^{(1)}[L]}
            {{\cal A}_1^{(2)}[L]}\,;\\
  \Delta_3^\text{FAPT}[L] 
    \!&\!=\!&\! \Delta_2^\text{FAPT}[L]
     - \frac{c_1^2\,\left({\cal D}^{2}+{\cal D}-1\right)\,{\cal A }_{\nu=3}^{(1)}[L]}
            {{\cal A}_1^{(2)}[L]}\,;\\
  \Delta_3^\text{PT}[L] 
    \!&\!=\!&\! 1
     - \frac{a_{(1)}[L]
           + c_1\,a_{(1)}^2[L]\,\ln\,a_{(1)}[L]
           + c_1^2\,a_{(1)}^3[L]
             \left(\ln^2 a_{(1)}[L] + \ln\,a_{(1)}[L] - 1\right)}
            {a_{(2)}[L]}\,.~~~
\end{eqnarray}
We see that relative accuracy of the 3-term approximation in FAPT
(see the left panel of Fig.\ \ref{fig:FAPT23_PT3}) is better 
than 2\% for $L\geq-2$.
In the same time, the right panel of Fig.\ \ref{fig:FAPT23_PT3}
demonstrates 
that relative accuracy of the same 3-term approximation in standard PT
even at $L\approx1$ is much higher --- about 10\%,
whereas in FAPT it is smaller than 1\%!
\begin{figure}[ht]
 \begin{minipage}{\textwidth}
  \centerline{\includegraphics[width=0.45\textwidth]{
              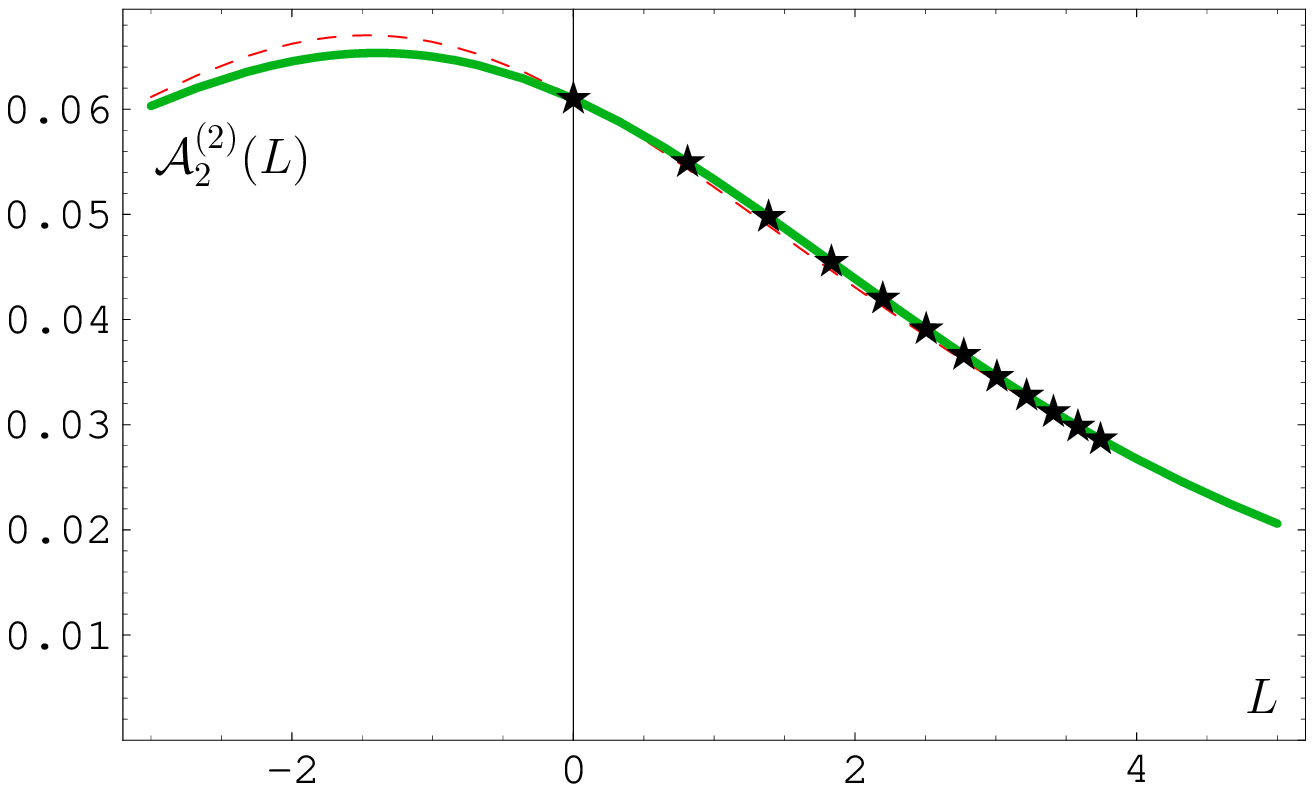}~~~%
              \includegraphics[width=0.45\textwidth]{
              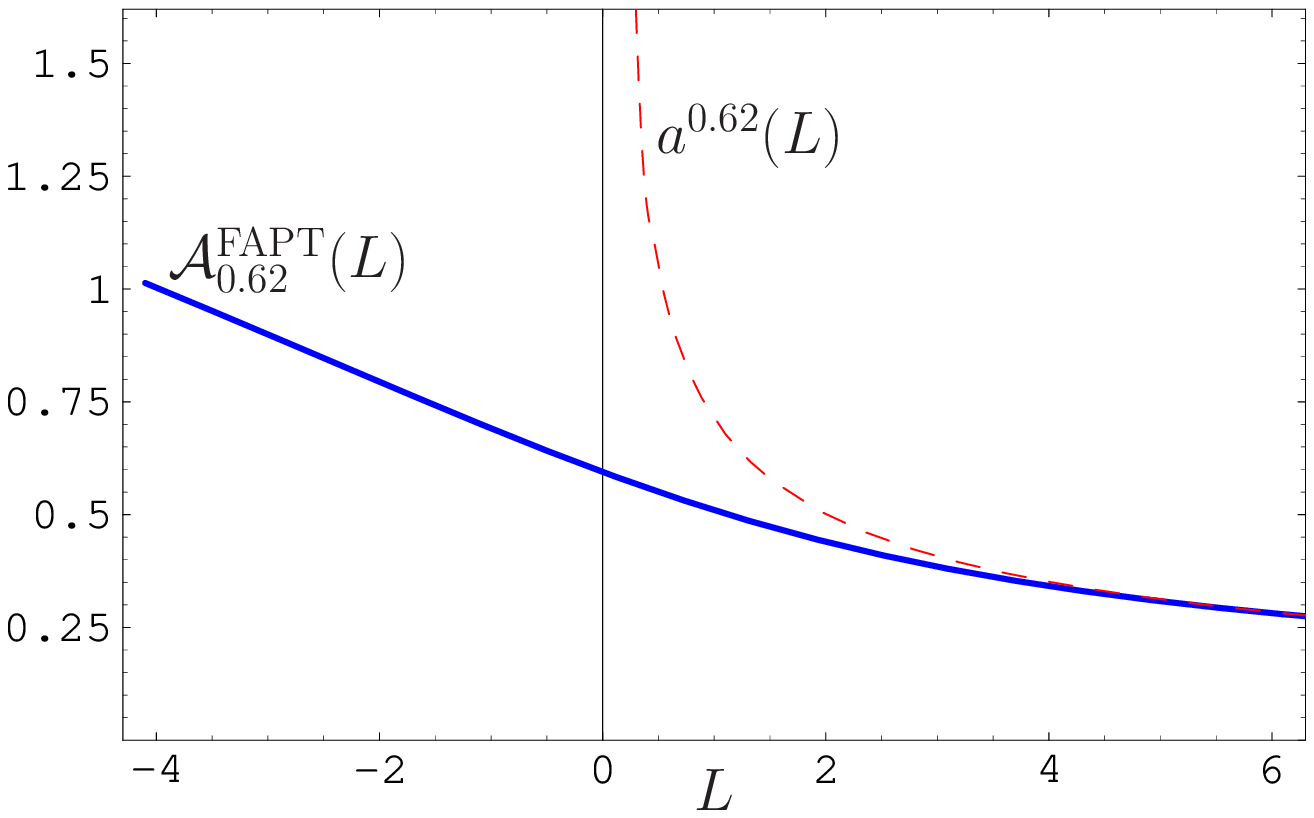}\vspace*{-1mm}}%
  \caption{\small {Left panel}: The solid line corresponds to ${\cal A}_2^{(2)}[L]$,
  computed analytically via Eq.\ (\ref{eq:A2_FAPT3});
  dashed line represents the result of a numerical integration,
  while stars correspond to the available numerical results 
  of Magradze in \cite{Mag00}.
  {Right panel}: The solid line represents ${\cal A}_{0.62}^{(2);\text{FAPT}}[L]$,
  computed analytically via Eq.\ (\ref{eq:A2_FAPT3}), while
  the dashed line stands for $a^{0.62}_{(2)}[L]$.
  \label{fig:Fapt-Pt.A2}}
\end{minipage}
\end{figure}
\\
We obtain the corresponding expansion for the two-loop coupling with
index $\nu$:
\begin{eqnarray}\!\!\!
 {\cal A}_{\nu}^{(2);\text{FAPT}}[L]
   &=& {\cal A }_{\nu}^{(1)}[L]
         + c_1\nu\,{\cal D}\,{\cal A }_{\nu+1}^{(1)}[L]
         + c_1^2\nu\left[\frac{\nu+1}{2}{\cal D}^{2}\!+\!{\cal D}\!-\!1\right]
              \!{\cal A }_{\nu+2}^{(1)}[L]
         + \ldots\,.~~~\label{eq:A2_FAPT3}
\end{eqnarray}
and display comparison of different results for ${\cal A}_2^{(2);\text{FAPT}}[L]$
on the left panel of Fig.\ \ref{fig:Fapt-Pt.A2}.
On the right panel of this figure 
we show  comparison of FAPT and standard QCD PT with respect
to the fractional index (power) of the coupling,
fixed at the value $\nu=0.62$,
corresponding to the evolution exponent of the second moment 
of the pion distribution amplitude.

In the Minkowskian region convergence of the FAPT expansion
for the two-loop coupling 
\begin{eqnarray}\!\!\!
 {\mathfrak A}_{2}^{(2);\text{FAPT}}[L]
  &=& {\mathfrak A }_{2}^{(1)}[L]
       + 2\,c_1\,{\cal D}\,{\mathfrak A }_{\nu=3}^{(1)}[L]
       + c_1^2\,\left[3\,{\cal D}^{2}+2\,{\cal D}-2\right]
         \!{\mathfrak A }_{\nu=4}^{(1)}[L]
       + \ldots\,.~~~
 \label{eq:MFAPT.U23}
\end{eqnarray} 
is also very good,
but in the vicinity of the point $L=0$ 
(Landau pole in the standard PT)
it is not so fast,
so that we need to take into account
$O(c_1^5)$-terms in order to reach 5\% level of accuracy,
for more details look in~\cite{BMS05}.

\begin{figure}[hb]
 \begin{minipage}{\textwidth}
  \centerline{\includegraphics[width=0.45\textwidth]{
              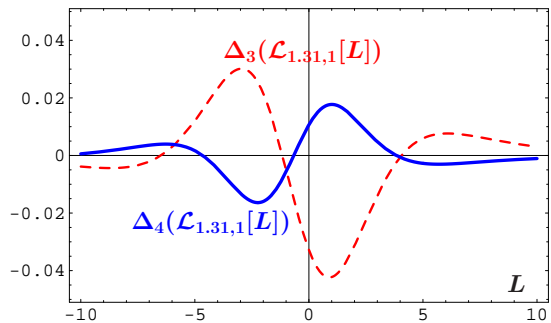}
   \vspace*{-1mm}}%
  \caption{\small The solid line corresponds to ${\cal A}_2^{(2)}[L]$,
  computed analytically via Eq.\ (\ref{eq:A2_FAPT3});
  dashed line represents the result of a numerical integration. 
  \label{fig:FAPT.Log.A_1.31}}
\end{minipage}
\end{figure}

Now we turn to the corresponding expansion 
for the two-loop coupling with logarithm:
\begin{eqnarray}
 \mathcal L_{\nu,1}^{(2)}[L]
   = \textbf{A}_\text{E}\left[\left(a_{(2)}\right)^\nu L\right]
   = \mathcal A_{\nu-1}^{(2)}
      + c_1\,\mathcal D\,\mathcal A_{\nu}^{(2)}
      - c_1^2\,\mathcal A_{\nu+1}^{(2)}
      + \frac{c_1^3}{2}\,\mathcal A_{\nu+2}^{(2)}
      - \frac{c_1^4}{3}\,\mathcal A_{\nu+3}^{(2)}
      + O\left(c_{1}^{5}\right)\,.
\end{eqnarray}
The exact spectral density can be easily found,
\begin{eqnarray}
 \rho^{(2)}_{\mathcal L_{\nu,1}}[L]
  = \frac{R_{(1)}[L]}
         {R_{(2)}^\nu[L]}
   \sin\left[\nu\varphi_{(2)}[L]
           - \varphi_{(1)}[L]
        \right]
\end{eqnarray}
with $R_{(1,2)}[L]$ 
and $\varphi_{(1,2)}[L]$ 
being inverse modula and phases of the corresponding 
one- and two-loop densities.  
In Fig.\ \ref{fig:FAPT.Log.A_1.31} we show the relative deviations
\begin{eqnarray}
 \Delta_{3,4}(\mathcal L_{1.31,1})
   = \frac{\mathcal L_{1.31,1}^{(1)}+O(c_1)+O(c_1^2)+O(c_1^3)+O(c_1^4)}
          {\mathcal L_{1.31,1}^{(2)}}
   -1\,.
\end{eqnarray}
Once again, we see that relative accuracy of the $O(c_1^4)$ approximation in FAPT
is better  than 2\% for all $L$
with the largest deviations in the vicinity of the point $L=0$.
In the Minkowski region convergence of the FAPT expansion
for the corresponding quantity $\mathfrak L_{\nu,1}^{(2)}[L]$
is also very good,
but not so fast as in the Euclidean case,
so that we need to take into account
$O(c_1^5)$-term in order to reach the same level of accuracy,
for more details see in~\cite{AB08}.

\subsection{Heavy-quark thresholds}
 \label{subsec:Global.FAPT}
Construction of FAPT with fixed number of quark flavors, $N_f$, 
is a two-step procedure: 
we start with the perturbative result $\left[a(Q^2)\right]^{\nu}$,
generate the spectral density 
$\rho_{\nu}(\sigma)=(1/\pi)\textbf{Im}\big[a^{\nu}(-\sigma)\big]$.,
and then obtain analytic couplings 
${\mathcal A}_{\nu}[L]$ and ${\mathfrak A}_{\nu}[L]$ via Eqs.\ (\ref{eq:A.U}).
Here $N_f$ is fixed and $N_f$-dependent coefficient $b_0(N_f)$ is factorized out.
We can proceed in the same manner for $N_f$-dependent quantities:
$\left[\alpha_s^{}(Q^2;N_f)\right]^{\nu}$ 
$\Rightarrow$ 
$\bar{\rho}_{\nu}(\sigma;N_f)=\bar{\rho}_{\nu}[L_\sigma;N_f]
 \equiv\rho_{\nu}(\sigma)/\beta_f^{\nu}$
$\Rightarrow$ 
$\bar{\mathcal A}_{\nu}^{}[L;N_f]$ and $\bar{\mathfrak A}_{\nu}^{}[L;N_f]$ ---
here $N_f$ is fixed, but not factorized out.

Global version of FAPT,
which takes into account heavy-quark thresholds,
is constructed along the same lines
but starting from global perturbative coupling
$\left[\alpha_s^{\,\text{\tiny glob}}(Q^2)\right]^{\nu}$,
being a continuous function of $Q^2$
due to choosing different values of QCD scales $\Lambda_f$,
corresponding to different values of $N_f$.
We illustrate here the case of only one heavy-quark threshold 
at $s=m_4^2$,
corresponding to the transition $N_f=3\to N_f=4$.
Then we obtain the discontinuous spectral density 
\begin{eqnarray}
 \label{eq:global_PT_Rho_4}
  \rho_n^\text{\tiny glob}(\sigma)
  =
  \rho_n^\text{\tiny glob}[L_\sigma]
   = \theta\left(L_\sigma<L_{4}\right)\,
       \bar{\rho}_n\left[L_\sigma;3\right]
    + \theta\left(L_{4}\leq L_\sigma\right)\,
       \bar{\rho}_n\left[L_\sigma+\lambda_4;4\right]\,,~~~
\end{eqnarray}
with $L_{\sigma}\equiv\ln\left(\sigma/\Lambda_3^2\right)$,
$L_{f}\equiv\ln\left(m_f^2/\Lambda_3^2\right)$
and
$\lambda_f\equiv\ln\left(\Lambda_3^2/\Lambda_f^2\right)$ for $f=4$,
which is expressed in terms of fixed-flavor spectral densities
with 3 and 4 flavors,
$\bar{\rho}_n[L;3]$ and $\bar{\rho}_n[L+\lambda_4;4]$.
However it generates the continuous Minkowskian coupling 
\begin{eqnarray}
 {\mathfrak A}_{\nu}^{\text{\tiny glob}}[L_s]
  \!&\!=\!&\! 
    \theta\left(L_s\!<\!L_4\right)
     \Bigl(\bar{{\mathfrak A}}_{\nu}^{}[L_s;3]
          -\bar{{\mathfrak A}}_{\nu}^{}[L_4;3]
          +\bar{{\mathfrak A}}_{\nu}^{}[L_4+\lambda_4;4] 
     \Bigr)
  \nonumber\\
  \!&\!+\!&\!
    \theta\left(L_4\!\leq\!L_s\right)\,
     \bar{{\mathfrak A}}_{\nu}^{}[L_s+\lambda_4;4]\,.
 \label{eq:An.U_nu.Glo.Expl}     
\end{eqnarray}
and the analytic Euclidean coupling (for more detail see in~\cite{AB08})
\begin{eqnarray}
 {\cal A}_{\nu}^{\text{\tiny glob}}[L]
  &=& \bar{{\cal A}}_{\nu}^{}[L+\lambda_4;4] 
    + \int\limits_{-\infty}^{L_4}\!
       \frac{\bar{\rho}_{\nu}^{}[L_\sigma;3]
            -\bar{\rho}_{\nu}^{}[L_\sigma+\lambda_{4};4]}
            {1+e^{L-L_\sigma}}\,
         dL_\sigma\,.
  \label{eq:Delta_f.A_nu}
\end{eqnarray} 

\begin{figure}[hb]
 \begin{minipage}{\textwidth}
  \centerline{\includegraphics[width=0.45\textwidth]{
              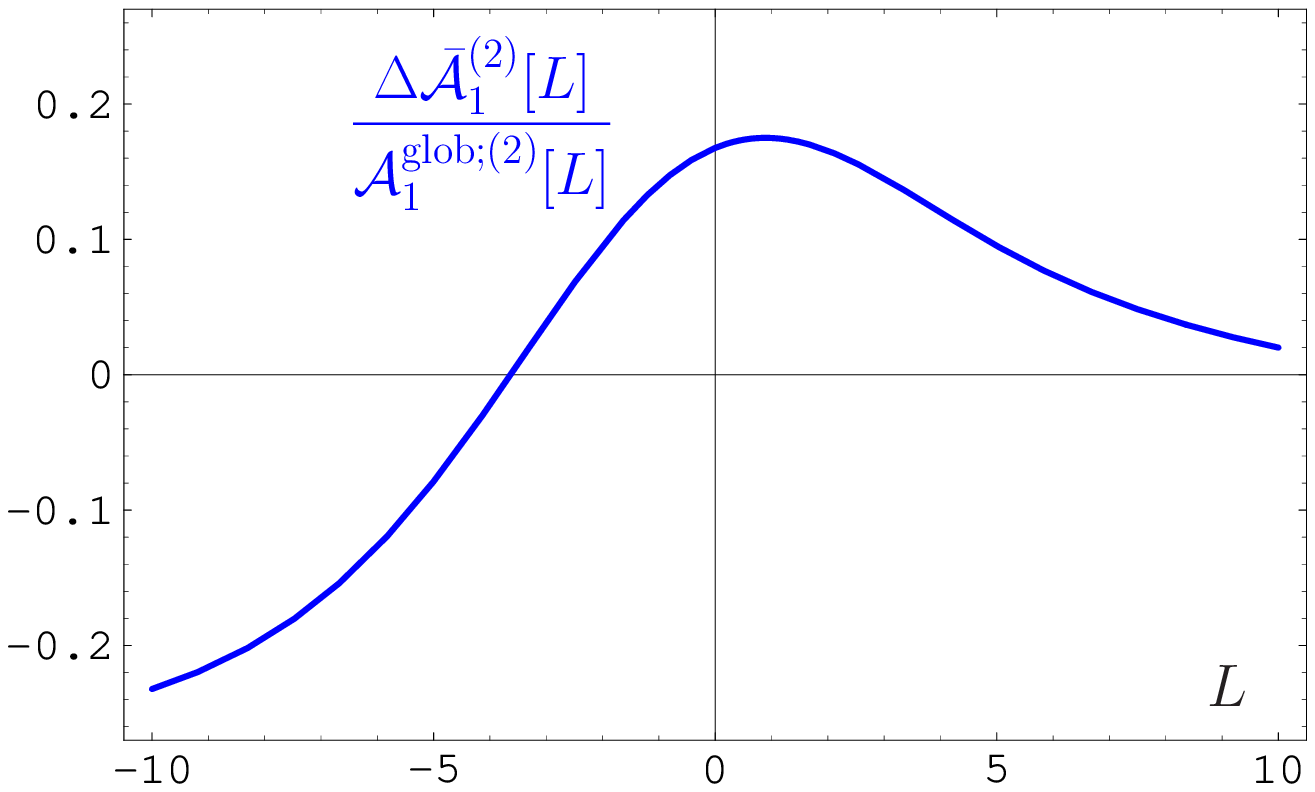}~~~%
              \includegraphics[width=0.45\textwidth]{
              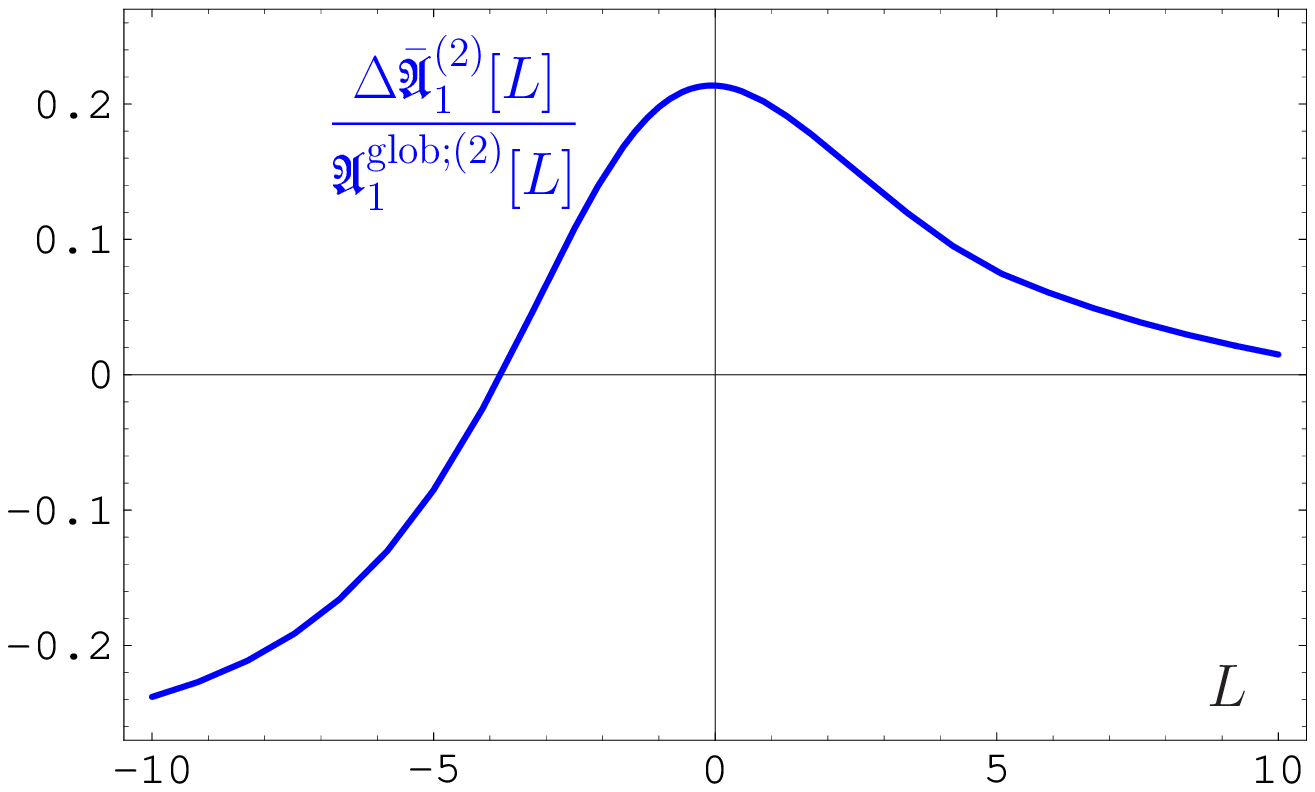}
   \vspace*{-1mm}}%
  \caption{\small Relative deviations of global coupling 
   with respect to fixed-$N_f$ coupling in APT.
   Solid lines correspond to the ratios 
   $\Delta\overline{\mathcal A}_{1}[L]/\mathcal A_{1}^\text{\tiny glob}[L]$
   (leftt panel) and 
   $\Delta\overline{\mathfrak A}_{1}[L]/\mathfrak A_{1}^\text{\tiny glob}[L]$
   (right panel). 
  \label{fig:Delta.Global.A_1}}
\end{minipage}
\end{figure}

We analyze now how important is the deviation of the global FAPT  
from fixed $N_f$ FAPT.
In the  Euclidean domain we may write
\begin{eqnarray}
 \mathcal A_{\nu}^\text{\tiny glob}[L]
  = \overline{\mathcal A}_{\nu}[L+\lambda_4;4]
  + \Delta\overline{\mathcal A}_{\nu}[L]
\end{eqnarray}
with
\begin{eqnarray}
 \Delta\overline{\mathcal A}_{\nu}[L]
  \equiv
   \int\limits_{-\infty}^{L_4}\!            
    \frac{\overline{\rho}_{\nu}\left[L_\sigma;3\right]-
          \overline{\rho}_{\nu}\left[L_\sigma\!+\!\lambda_4;4\right]}
         {1+e^{L-L_\sigma}}\,dL_\sigma\,,
\end{eqnarray}
whereas in the Minkowski domain ---
\begin{eqnarray}
 {\mathfrak A}_{\nu}^\text{\tiny glob}[L]
  = \overline{\mathfrak A}_{\nu}[L\!+\!\lambda_4;4]
  + \Delta\overline{\mathfrak A}_{\nu}[L]
\end{eqnarray}
with
\begin{eqnarray}
 \Delta\overline{\mathfrak A}_{\nu}[L]
  \equiv \theta\left(L<L_{4}\right)
  \!\!&\!\!\Big[\!\!&\!\!
    \overline{\mathfrak A}_{\nu}[L;3]
  - \overline{\mathfrak A}_{\nu}[L_4;3]
  + \overline{\mathfrak A}_{\nu}[L_4\!+\!\lambda_4;4]
  - \overline{\mathfrak A}_{\nu}[L\!+\!\lambda_4;4]
          \Big]
\end{eqnarray}
In Fig.\ \ref{fig:Delta.Global.A_1} we show the relative values 
of deviations
in Euclidean and Minkowski domains.
We see that in both domains deviations vary from $-20\%$ at large values of 
$-L\approx10$,
in the vicinity of $L\approx-5$ they go through the zero,
then grow up to the value $+20\%$ which is reached at $L\approx0$, 
and then tends to 0 at $L\to\infty$.

\section{Electromagnetic pion form factor at NLO}
 \label{sec:Pion.FF}
The scaled hard-scattering amplitude truncated at the next-to-leading order 
(NLO) 
and evaluated at renormalization scale $\mu_{R}^2=\lambda_{R} Q^2$ 
reads~
\cite{FGOC81,DR81,BT87,MNP99a}
\begin{eqnarray}
  T^\text{NLO}_\text{H}\left(x ,y;\mu_{F}^2,Q^2\right)
  = \frac{\alpha_s\left(\lambda_{R}Q^2\right)}{Q^2}\,
     t_\text{H}^{(0)}(x,y)
  + \frac{\alpha_s^2\left(\lambda_{R}Q^2\right)}{4\pi\,Q^2}\,
      t_\text{H}^{(1)}(x,y;{\mu_{F}^2}/{Q^2})~~~\label{eq:T_Hard_NLO}  
\end{eqnarray}
with shorthand notation ($\bar{x}\equiv1-x$)
\begin{eqnarray}
 t_\text{H}^{(1)}(x,y;{\mu_{F}^2}/{Q^2})
  = C_{F}\,t_{\text{H}}^{(0)}\left(x,y\right)\,
          2\Big(3+\ln\,(\bar{x}\bar{y})\Big)
      \ln\frac{Q^2}{\mu_{F}^2}
        + b_0\,t_\text{H}^{(1,\beta)}\left(x,y;\lambda_{R}\right)
        + t_\text{H}^{(\text{FG})}(x,y)\,.~
\end{eqnarray}
The leading twist-2 pion distribution amplitude (DA)~\cite{Rad77} 
at normalization scale $\mu_F^2$ 
is given by~\cite{ER80}
\begin{eqnarray}
 \varphi_\pi(x,\mu_F^2)
  =  6\,x\,(1-x)
      \left[ 1
         + a_2(\mu_F^2) \, C_2^{3/2}(2 x -1)
         + a_4(\mu_F^2) \, C_4^{3/2}(2 x -1) 
         + \ldots\,
      \right]\,.
\end{eqnarray}
All {nonperturbative} information is encapsulated in Gegenbauer coefficients 
$a_n(\mu^2_F)$.

To obtain factorized part of pion form factor (FF) one needs to convolute
the pion DA with the hard-scattering amplitude:
\begin{eqnarray}
 F_\pi^\text{Fact}(Q^2) 
  = \varphi_\pi(x;\mu_{F}^2)\mathop{\otimes}\limits_{x}
      T^\text{NLO}_\text{H}\left(x,y;\mu_{F}^2,Q^2\right)
         \mathop{\otimes}\limits_{y}\varphi_\pi(y;\mu_{F}^2)\,.
\end{eqnarray}
In order to obtain the analytic expression for the pion FF at NLO 
in~\cite{SSK99} the so-called ``Naive Analytization'' 
has been suggested.
It uses analytic image only for coupling itself, ${\cal A}_{1}^{(2)}$,
but not for its powers.
In contrast and in full accord with the APT ideology 
the receipt of ``Maximal Analytization'' has been proposed
recently in~\cite{BPSS04}. 
The corresponding expressions for the analytized hard amplitudes
read as follows:
\begin{eqnarray}
 \left[Q^2 T_\text{H}\left(x,y,Q^2\right)
 \right]_\text{Nai-An} 
 &=& {\cal A}_{1}^{(2)}(\lambda_{R} Q^2)\,t_\text{H}^{(0)}(x,y)
 + \frac{\left({\cal A}_{1}^{(2)}(\lambda_{R} Q^2)\right)^2}{4\pi}\,
    t_\text{H}^{(1)}\left(x,y;\lambda_{R},\frac{\mu_{F}^2}{Q^2}\right)\,;\\
 \left[Q^2 T_\text{H}\left(x,y,Q^2\right)
 \right]_\text{Max-An}
 &=& {\cal A}_{1}^{(2)}(\lambda_{R} Q^2)\,t_\text{H}^{(0)}(x,y)
    + \frac{{\cal A}_{2}^{(2)}(\lambda_{R} Q^2)}{4\pi}\,
       t_\text{H}^{(1)}\left(x,y;\lambda_{R},\frac{\mu_{F}^2}{Q^2}\right)\,.
\end{eqnarray}
\begin{figure}[h]
 \begin{minipage}{\textwidth}
  \centerline{\includegraphics[width=0.325\textwidth]{
              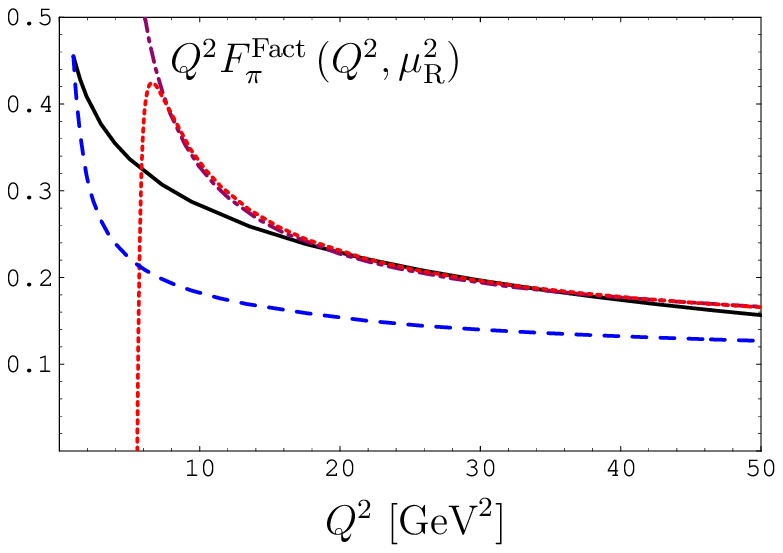}~
              \includegraphics[width=0.325\textwidth]{
              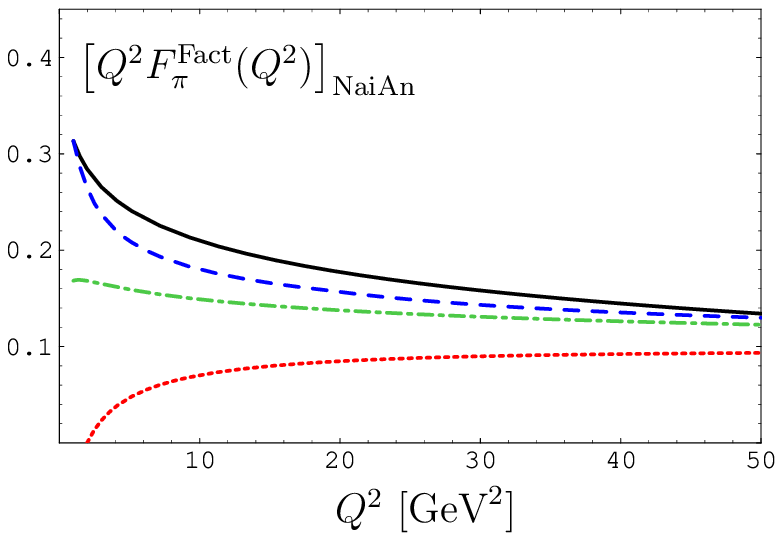}~%
              \includegraphics[width=0.325\textwidth]{
              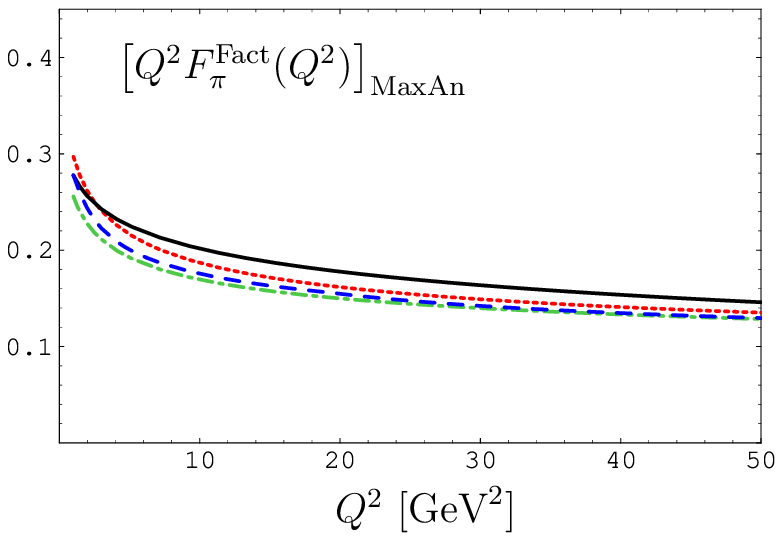}}
  \caption{\small {Left panel}: Factorized pion FF in the standard 
   $\overline{\strut\text{MS}}$ scheme.
   The solid line corresponds to the scale setting $\mu_\text{R}^2=1$~GeV$^2$, 
   dashed line --- to $\mu_\text{R}^2=Q^2$, dotted lines --- to the principle of minimal 
   sensitivity, whereas dash-dotted lines --- to the principle of fastest 
   apparent convergency.   
   {Central panel}: Factorized pion FF in the ``Naive Analytization''.
   {Right panel}: Factorized pion FF in the ``Maximal Analytization''.
   On both panels solid lines correspond to the scale setting $\mu_\text{R}^2=1$~GeV$^2$, 
   dashed lines --- to $\mu_\text{R}^2=Q^2$, dotted lines --- 
   to the BLM~\cite{BLM83} prescription,
   whereas dash-dotted lines --- to the $\alpha_\text{V}$-scheme.
     \label{fig:PionFF_Nai_Max}}
\end{minipage}
\end{figure}
In Fig.\ \ref{fig:PionFF_Nai_Max} we show the predictions for
the factorized pion FF in the standard pQCD and 
in ``Naive'' and in the ``Maximal Analytization'' approaches. 
We see that in the ``Maximal Analytization'' approach
the obtained results are practically insensitive 
to the renormalization scheme and scale-setting choice
(already at the NLO level).

We show also the graphics for the whole pion FF,
obtained in APT with the ``Maximally Analytic'' procedure
using the Ward identity to match the non-factorized and factorized parts
of the pion FF, see the right panel of Fig.\ \ref{fig:PionFF_Nai_Max}.
The green strip in this figure contains both nonperturbative uncertainties 
from nonlocal QCD sum rules~\cite{BMS01,BP06,AB06parus} 
and renormalization scheme and scale ambiguities 
at the level of the NLO accuracy.

It is interesting to note here that the FAPT approach,
used in~\cite{BKS05} for analytization of the $\ln(Q^2/\mu_{F}^2)$-terms 
in the hard amplitude (\ref{eq:T_Hard_NLO}),
diminishes also the dependence on the factorization scale setting 
in the interval $\mu_{F}^2=1-50$~GeV$^2$.

\begin{figure}[h]
 \begin{minipage}{\textwidth}
  \centerline{\includegraphics[width=0.45\textwidth]{
              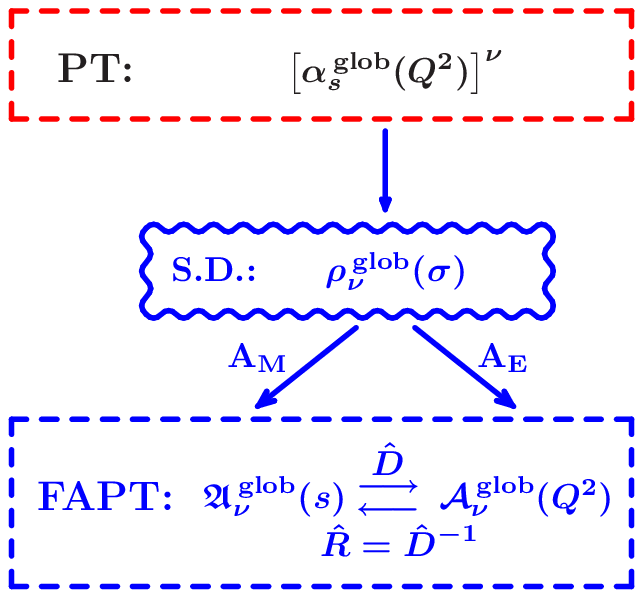}~~~
              \includegraphics[width=0.45\textwidth]{
              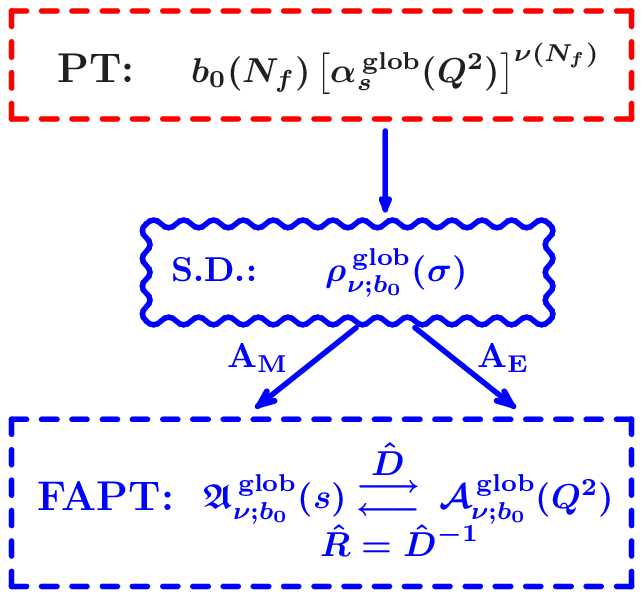}}
  \caption{\small Schemes of ``analytization'' of $N_f$-dependent quantities.
   On the left panel the starting object to analytize 
   is $[\alpha_s^\text{\tiny glob}(Q^2)]^\nu$,
   whereas on the right panel --- $b_0(N_f)[\alpha_s^\text{\tiny glob}(Q^2)]^\nu$.
     \label{fig:Sceme.APT.Glob.b0}}
\end{minipage}
\end{figure}

This conclusion starts to be even more pronounced 
in the complete FAPT analysis of the factorized pion FF.
Indeed, if we put $\mu_\text{F}^2=Q^2$ then we obtain 
in the pion FF convolutions with $\varphi_\pi(x,Q^2)$
which contains ERBL evolution factors 
$a_{2n}(\mu_{0}^2)\left[\alpha_s(Q^2)/\alpha_s(\mu_{0}^2)\right]^{\nu_{2n}}$
with $\nu_{2n}(N_f)=\gamma_{0}(2n)/(2b_0(N_f)$.
Numerically $\nu_{2}=0.62-0.72$ and 
$\nu_{4}=0.90-1.06$.
In  $T_\text{H}^\text{NLO}\left(x,y,Q^2\right)$ 
we have three types of contributions:
$\alpha_s(Q^2)$, $\alpha_s^2(Q^2)$, and $b_0(N_f)\alpha_s^2(Q^2)$.
The scheme of ``analytization'' of $N_f$-dependent quantity
is the same as in the case of the global approach,
see in Fig.\ \ref{fig:Sceme.APT.Glob.b0}.
Indeed, 
we start with $b_0(N_f)\alpha_s^\nu(Q^2)$,
obtain corresponding spectral density
$\rho^\text{\tiny glob}_{\nu;b_0}(\sigma)$,
being in case of the single heavy-quark threshold
of the type
\begin{eqnarray}
 \label{eq:global_PT_Rho-b0_4}
  \rho^\text{\tiny glob}_{\nu;b_0}(\sigma)
  =
  \rho^\text{\tiny glob}_{\nu;b_0}[L_\sigma]
   = \theta\left(L_\sigma<L_{4}\right)\,
       b_0(3)\,\bar{\rho}_n\left[L_\sigma;3\right]
    + \theta\left(L_{4}\leq L_\sigma\right)\,
       b_0(4)\,\bar{\rho}_n\left[L_\sigma+\lambda_4;4\right]\,,~~~
\end{eqnarray}
and then construct corresponding analytic image
$\mathcal A^\text{\tiny glob}_{\nu;b_0}[L]$,
using integral representations (\ref{eq:AU_n}).

\begin{figure}[h]
 \begin{minipage}{\textwidth}
  \centerline{\includegraphics[width=0.5\textwidth]{
              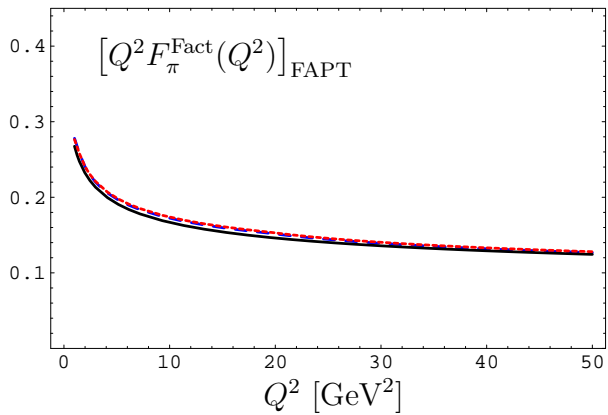}}
  \caption{\small Factorized pion FF in the complete FAPT approach.
   The solid line corresponds to the factorization scale $\mu_\text{F}^2=Q^2$, 
   dashed line --- to $\mu_\text{F}^2=6$~GeV$^2$, 
   whereas dotted line --- to the``Maximal Analytization'' with $\mu_\text{F}^2=Q^2$.
     \label{fig:PionFF_Max_Complete}}
\end{minipage}
\end{figure}

In Fig.\ \ref{fig:PionFF_Max_Complete} we show the predictions for
the factorized pion FF in the complete FAPT approaches 
with different settings for $\mu_\text{F}$
(see figure caption for details)
and for comparison --- in the ``Maximal Analytization'' scheme 
with $\mu_\text{F}^2=Q^2$.
We see that both approaches
produce practically the same results.

We conclude this section with 
the note 
that in the  Euclidean domain taking thresholds and $N_f$-dependence 
of coefficients 
into account generates tiny correction!
The main advantage of the complete FAPT approach: 
There are no problems with thresholds.
Bonus: Pion FF automatically appears to be analytic function 
out of the Minkowski cut~\cite{AB08}.
Note here
that in this case one should use in the Minkowski region 
not the coupling
$\mathfrak A_\nu(s)$,
but instead the analytic image of the  Euclidean coupling
on the lower side of the cut,
namely, 
$\mathcal A_\nu(-s-i0)=\textbf{Re}\,\mathcal A_\nu(s)+i\pi\rho_\nu(s)$.
In contrast to the real coupling $\mathfrak A_\nu(s)$,
this quantity naturally generates 
an imaginary part of the pion form factor,
which is measured in $e^+e^-\to\pi^+\pi^-$ reactions.

\section{Higgs boson decay into $b\bar{b}$ pair in FAPT}
 \label{sec:Higgs.Decay}
In this section we analyze the Higgs boson decay to a $\bar{b}b$ pair.
For the decay width we have  
\begin{eqnarray}
 \Gamma_{\text{H}\to\bar{b}b}(M_\text{H})
  = \frac{G_F}{4\sqrt{2}\pi}\,
     M_\text{H}\,
      \widetilde{R}_\text{\tiny S}(M_\text{H}^2)
  ~~~\text{with}~~~
  \widetilde{R}_\text{\tiny S}(M_\text{H}^2)
  \equiv m^2_{b}(M_\text{H}^2)\,R_\text{\tiny S}(M_\text{H}^2)
 \label{eq:Higgs.decay.rate}
\end{eqnarray}
and
$R_\text{\tiny S}(s)$ 
is the $R$-ratio for the scalar correlator,
see for details in~\cite{BMS06,BCK05}.
The running mass $m(Q^2)$ is described in the three-loop approximation
by the renormgroup equation~\cite{AB08}
\begin{eqnarray}
 m_{(3)}^2(Q^2)
  = \hat{m}_{(3)}^2
     \left[\alpha_{s}(Q^2)\right]^{\nu_0}
      \left[1 - \delta_{21}\,\alpha_s\right]^{\nu_{21}}\,
       \left[1 + \delta_{22}\,\alpha_s\right]^{\nu_{22}}\,,
\end{eqnarray}
with RG-invariant mass $\hat{m}_{(3)}^2$ 
(for $b$-quark $\hat{m}_{b;(3)}\approx 8~\text{GeV}$)
and $\nu_0=1.04$, $\delta_{21}=0.672$, $\nu_{21}=-0.743$, 
                  $\delta_{22}=0.029$, and $\nu_{22}=8.59$
at $N_f=5$.

In the standard PT direct multi-loop calculations 
are usually performed 
in the Euclidean region for the corresponding Adler function 
$D_\text{S}=3\,\left[1+\sum_{n>0}d_n\,\alpha_{s}^n(Q^2)\right]$, 
where QCD perturbation theory works.    
Functions $D_\text{S}$ and 
$R_\text{S}(s)=3\,\left[1+\sum_{n>0}r_{n}~\alpha_{s}^n(s)\right]$
can be related to each other 
via a dispersion relation 
without any reference to perturbation theory.
This generates relations between $r_n$ and $d_n$ 
coefficients,
involving the famous $\pi^2$-terms
due to integral transformation of $\ln^{k}(Q^2/\mu^2)$ 
in $d_{n}$:
$$
 \mathfrak A_{-2}[L]=L^2-\frac{\pi^2}{3}, ~~~
 \mathfrak A_{-3}[L]=L\left(L^2-\pi^2\right),~\ldots\,.
$$ 
The influence of these $\pi^2$-terms can be substantial, 
as has been shown by Baikov\textit{et al.}~\cite{BCK05}.

\subsection{Comparing different approaches to calculate 
 $\Gamma_{\text{H}\to\bar{b}b}(M_\text{H})$}
\label{subsec:various-Rs}
\begin{figure}[h]
 \centerline{\includegraphics[width=0.47\textwidth]{
             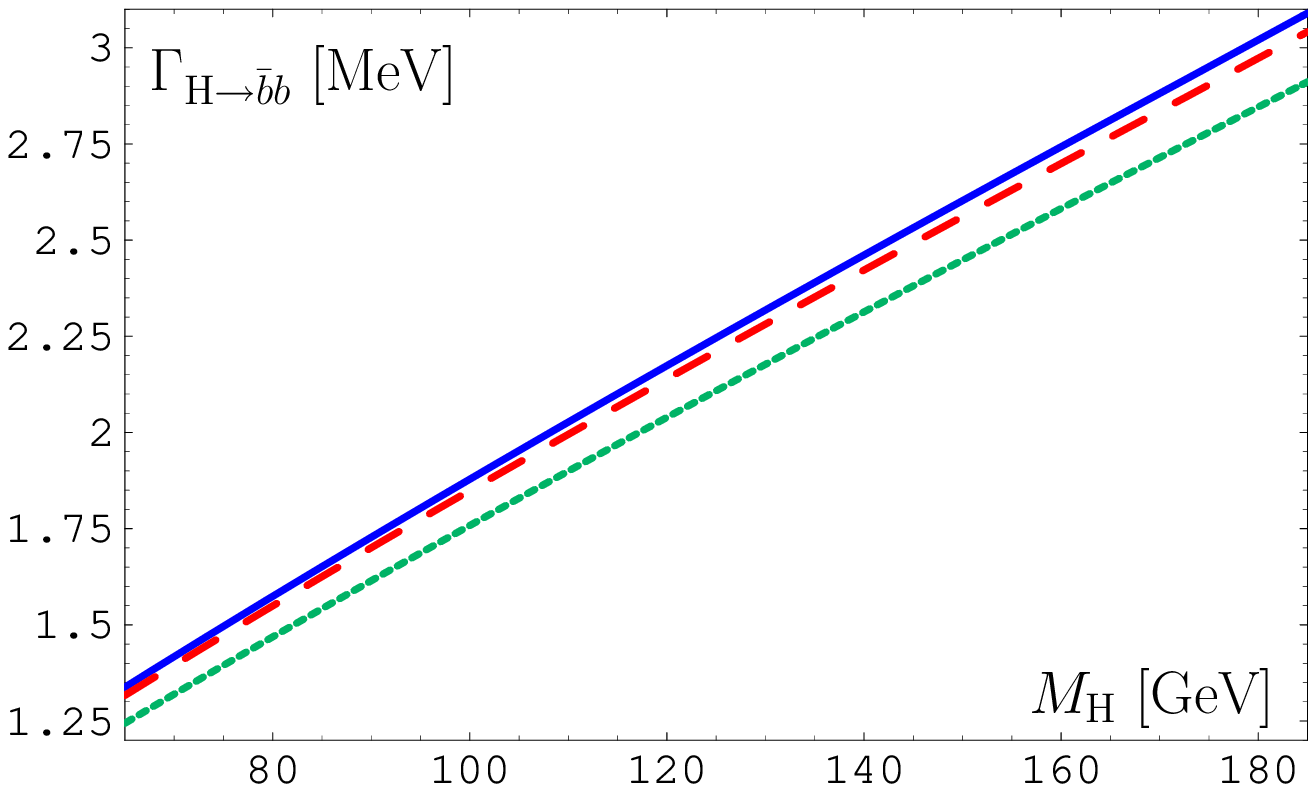}~
             \includegraphics[width=0.47\textwidth]{
             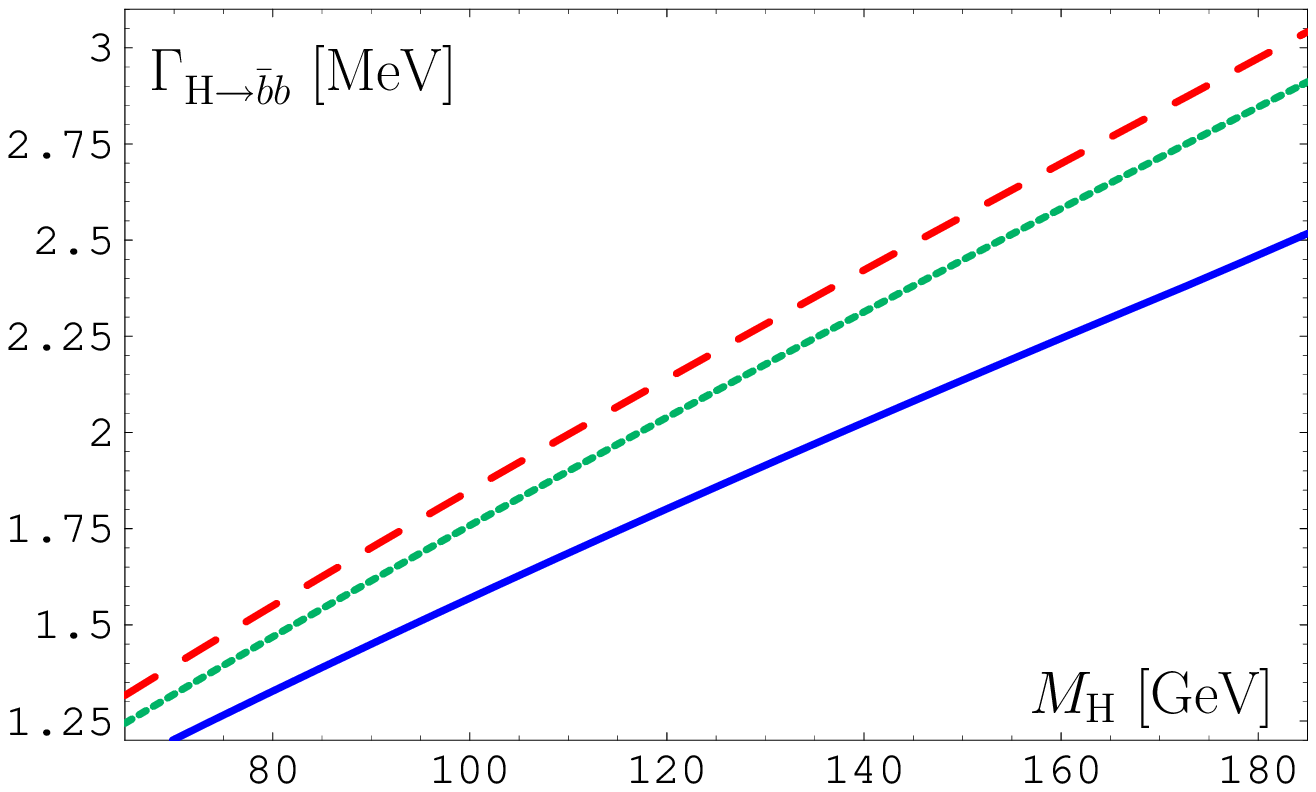}}
   \caption{\small Results of calculations of $\Gamma_{\text{H}\to\bar{b}b}(M^2_\text{H})$
    in different approaches. On both panels dashed (red) lines display the BCK results
    and dotted (green) lines --- the BKM results.
    Solid line on the left panel corresponds to the results obtained in the FAPT approach
    with freezed coefficients $d_n=d_n(N_f=5)$,
    whereas on the right panel --- to the results of the complete FAPT approach.
    \label{fig:R_S}}
\end{figure}

We compare now the results of different approaches to calculate 
$\widetilde{R}_\text{S}(M_\text{H}^2)$.
\begin{itemize}
 \item Baikov~\textit{et al.} (BCK)~\cite{BCK05} 
 used standard QCD PT at the $O(a_\text{s}^{4})$-order
 with $\Lambda_{N_f=5}^{(4)}=231$~MeV:
\begin{eqnarray}
 \widetilde{R}_{\text{S}}^{(l=4)\text{BCK}}(s)
  &=& 3 m_{(l=4)}^2(s)
        \left[1 + \sum_{n\geq 1}^{4} r_{n}(5)
                   \left(\frac{\alpha_\text{s}^{(l=4)}}{\pi}\right)^{n}
        \right]\,.
 \label{eq:R-Che}
\end{eqnarray}

 \item Broadhurst \textit{et~al.} (BKM)~\cite{BKM01} 
 in the approach of the so-called ``Naive Non-Abelianization'' (NNA) 
 used the ``contour-improved'' optimization of power expansion.
 Their formulas are very closed to the one-loop FAPT
 (see more detailed discussion in~\cite{BMS06,KK08}):
\begin{eqnarray}
 \widetilde{R}_{\text{S}}^{(l=1)\text{FAPT}}(s)
  = 3\,\hat{m}_{(l=1)}^2\,
      \left[{\mathfrak A}_{\nu_{0}}^{(1);\text{\tiny glob}}(s)
          + \sum_{n\geq1}^{4} d_{n}(5)
             \frac{{\mathfrak A}_{n+\nu_{0}}^{(1);\text{\tiny glob}}(s)}
                  {\pi^{n}}
      \right]\,.
\label{eq:R.MFAPT.1L}
\end{eqnarray}
 We show the prediction obtained in this approach
 with using $\Lambda_{N_f=5}^{(1);Z}=111$~MeV
 (extracted from condition ${\mathfrak A}_{1}^{(1);\text{\tiny glob}}(m_Z^2)=0.120$)
 by the dotted (green) line in Fig.~\ref{fig:R_S}. 
 It appears to be lower than the standard PT result (dashed red line) 
 by $\approx8$\%.
\end{itemize}

\begin{itemize} 
\item In the FAPT approach with ``freezed'' at the value $d_n(5)$ 
 coefficients $d_n(N_f)$ (as also has been done in the BCK and BKM approaches)
 we define $\mathfrak B_{n+\nu_0}(s)$
 as  the analytic images of 
 $\alpha_{s}^{n+\nu_0}(Q^2)[1+\delta_1\,\alpha_{s}(Q^2)]^{\nu_1}$
 in the Minkowski region 
 and obtain 
\begin{eqnarray}
 \widetilde{R}_\text{S}^{(3)\text{FAPT(5)}}(s)
   =  3\hat{m}_b^2
      \left[{\mathfrak B}_{\nu_{0}}^{(3);\textbf{\tiny glob}}(s)
          + \sum_{n\geq1}^{3} d_{n}(5)
             \frac{{\mathfrak B}_{n+\nu_{0}}^{(3);\textbf{\tiny glob}}(s)}
                   {\pi^{n}}
      \right]\,.
\end{eqnarray}

 The corresponding predictions for the value $\Lambda_{N_f=5}^{(3)}=220$~MeV
 (being in accord with the last analysis of precision electroweak data 
  by the LEP Collaboration~\cite{LEP07} and normalized 
  to ${\mathfrak A}_{1}^{(3);\text{\tiny glob}}(m_Z^2)=0.118$)
  are represented in the left panel of Fig.~\ref{fig:R_S} by the solid (blue) line.
 We see that it is very close to the BCK curve 
 (difference is of the order of 1.5\%, 
 and as compared with three-loop BCK result --- even of the order of 1\%).

\item In the complete FAPT approach with the complete analytization 
 of $N_f$-dependencies we obtain 
\begin{eqnarray}
 \widetilde{R}_\text{S}^{(3)\text{FAPT}}(s)
   =  3\hat{m}_b^2
      \left[\mathfrak B_{\nu_{0}}^{(3);\text{\tiny glob}}(s)
          + \sum_{n\geq1}^{l}
             \frac{\mathfrak B_{n+\nu_{0};d_n}^{(3);\text{\tiny glob}}(s)}
                  {\pi^{n}}
     \right]\,,
\end{eqnarray}
 where the analytic images $\mathfrak B_{n+\nu_0;d_n}(s)$
 absorb all $N_f$-dependence of $d_n$ coefficients.
 Here $\Lambda_{N_f=5}^{(3)}=227$~MeV in order to match 
 $R(m_Z^2)$ value by the LEP Collaboration~\cite{LEP07}
 in the complete FAPT approach:
 for this reason ${\mathfrak A}_{1}^{(3);\text{\tiny glob}}(m_Z^2)=0.119$
 and differs slightly from the ``freezed'' FAPT value 
 ${\mathfrak A}_{1}^{(3);\text{\tiny glob}}(m_Z^2)=0.118$.
 The results of this approach are shown in the right panel of Fig.\ \ref{fig:R_S}
 by the solid (blue) line
 and appear to be smaller than the ``freezed'' results.
 This difference varies from 14\% (at $M_\text{H}=50$~GeV)
 to 18\% (at $M_\text{H}=150$~GeV).
\end{itemize}
We can conclude from this comparison
that the standard pQCD power series 
and the FAPT non-power series expansions
appears to be quite close to each other in the region $L\gg1$
in the scenario with ``freezed'' coefficients $d_n(N_f)$,
corresponding to the value $N_f=5$.
The complete FAPT setup,
which bears in mind the whole $N_f$ dependence 
of the perturbative results,
naturally generates another, in our case smaller,
result.
\begin{table}[ht]\vspace*{-3mm}
 \caption{The effective coefficients $d_{n}^\text{eff}[L]$, 
  see~(\ref{eq:d.n.eff}), at $L=11-13$. 
  In two last columns we show the corresponding values 
  of $\widetilde{R}_\text{S}^{(3)}[L]$
  in cases of the ``freezed'' and the complete FAPT. \label{tab:d.n.eff}}
\begin{center}
 \begin{tabular}{ccccc||cc}
\hline 
      &~~$n=0$~~&~~$n=1$~~&~~$n=2$~~&~~$n=3\vphantom{^\big|_\big|}$~~
                                              &~~$\widetilde{R}_\text{S}^{(3);\text{FAPT(5)}}[L]$  
                                                        &~~$\widetilde{R}_\text{S}^{(3);\text{FAPT}}[L]$
\\ \hline
 $d_{n}(N_f=5)$
      & 1.00    & 5.67    & 42.0    & 353     &  --     &  --   
\\
 $d_{n}^\text{eff}[L=10]$  
      & 0.85    & 4.93    & 36.4    & 296     & 32.44   & 27.78
\\
 $d_{n}^\text{eff}[L=11]$
      & 0.83    & 4.80    & 34.9    & 276     & 29.11   & 24.41
\\
 $d_{n}^\text{eff}[L=12]$  
      & 0.82    & 4.66    & 33.2    & 252     & 26.38   & 21.64
\\ \hline      
 \end{tabular}
\end{center}
\end{table}
This is not a surprise,
because such an analytization approach effectively 
average coefficients $d_n(N_f)$ in the regions
with different $N_f$ values,
resulting in reductions of their values 
and, hence, of the whole non-power series sum.
In order to make this effect more transparent 
we define effective coefficients,
corresponding to the complete FAPT series
$\widetilde{R}_\text{S}^{(3);\text{FAPT}}[L]$
as follows:
\begin{eqnarray}
 \label{eq:d.n.eff}
  d_{n}^\text{eff}[L] 
   &=& \frac{{\mathfrak B}_{n+\nu_{0};d_n}^{(3);\text{\tiny glob}}[L]}
            {{\mathfrak B}_{n+\nu_{0}}^{(3);\text{\tiny glob}}[L]}\,.               
\end{eqnarray}
In Table~\ref{tab:d.n.eff}, 
we show these effective coefficients:
one can immediately realize
that they indeed diminished by 16 to 18\%.
Physically this effect of the complete analytization 
of $\widetilde{R}_\text{S}(s)$
corresponds to taking into account contributions of loops 
with $t$-quarks even in the region $\sqrt{s}\leq 175$~GeV, 
where in the standard pQCD in the $\overline{\text{MS}}$-scheme 
only quarks $u$, $d$, $s$, $c$, and $b$ contribute.
We see that this effect is sizable.

\section{Concluding Remarks}
We conclude with the following resume:
\begin{itemize}
  \item The implementation of the analyticity concept
    (the dispersion relations) from the level of the coupling 
     and its powers to the level of QCD amplitudes as a whole
     generates extension of the APT to FAPT.
  \item  The rules how to apply FAPT at the two- and three-loop levels
     are formulated.
  \item  The convergence of the perturbative expansion 
     is significantly improved when using non-power FAPT expansion;
  \item  We formulate the rules how to account for heavy-quark thresholds
         in FAPT.
  \item  As an additional advantage we obtain the minimal sensitivity 
         to both the renormalization and factorization scale setting,
         revealed on the example of the pion electromagnetic form factor.
         Threshold problem (how to fix the value of $b_0(N_f)$ 
         in the $O(\alpha_s^2)$-term) resolved.
         The result of the complete FAPT prescription appears 
         to be very close to the result of  the 
         Brodsky--Lepage--Mackenzie prescription,
         see in Figs.\ \ref{fig:PionFF_Nai_Max}
         and \ref{fig:PionFF_Max_Complete}.
  \item  In application to decay $\text{H}^0\to b\bar{b}$
         we revealed that the complete FAPT analytization 
         prescription reduces the results by $\approx16$\%
         as compared with the ``freezed'' FAPT ones.
\end{itemize}

The comparison of FAPT in Euclidean and Minkowski regions,
done in Sections 2 and 3,
shows us close analogy between the two regions---the effect,
named by Shirkov and Solovtsov 
as effect of ``distorting mirror''.
But the results of the complete FAPT applications to 
the analysis of the pion form factor in the Euclidean domain
(Section~\ref{sec:Pion.FF})
and of the Higgs boson decay width in the Minkowski domain
(Section~\ref{sec:Higgs.Decay})
are formally quite different:
For the pion form factor case the complete FAPT prescription 
generates approximately the same result 
as standard FAPT with ``freezed'' $N_f$-dependent terms~\cite{BPSS04,BKS05},
see in Fig.~\ref{fig:PionFF_Max_Complete},
whereas for the Higgs boson decay 
this approach generates reduction of the order of 16\%.  

In order to explain these difference
we analyzed the Adler function $D_\text{S}(Q^2)$ 
with the same perturbative coefficients $d_n(N_f)$.
Then we revealed that the magnitude of the corresponding reduction 
of the complete FAPT results in the  Euclidean region
in comparison with the ``freezed'' FAPT ones
is of the same order (18\%) as in the Minkowski case.
This demonstration proves 
that the magnitude of this reduction is related 
with the strong dependence of perturbative coefficients 
$d_n(N_f)$ on $N_f$
and not with specific character of analytization procedure 
in the Minkowski region.

\section*{Acknowledgments:}
I would like to thank  my colleagues and coauthors 
Sergey Mikhailov and Nico Stefanis for helpful discussions and support,
and Konstantin Chetyrkin, Andrei Kataev, and Alexei Pivovarov 
for stimulating discussions 
during the International Seminar ``Quarks-2008''
(Sergiev Posad, Russia, 23--29 May, 2008).
This investigation was supported in part 
by the Deutsche Forschungsgemeinschaft
(Projects DFG 436 RUS 113/881/0),
the Heisenberg--Landau Programme, grant 2008, 
the Russian Foundation for Fundamental Research, 
grants No.\ ü~06-02-16215, 07-02-91557, and 08-01-00686,
and the BRFBR--JINR Cooperation Programme, contract No.\ F06D-002.


\begin{thebibliography}{10}

\bibitem{JS95-349}
 H.~F. Jones and I.~L. Solovtsov,
  Phys. Lett. \textbf{B349},  519  (1995).

\bibitem{JS95-357}
 H.~F. Jones, I.~L. Solovtsov, and O.~P. Solovtsova,
  Phys. Lett. \textbf{B357},  441  (1995).

\bibitem{SS96}
 D.~V. Shirkov and I.~L. Solovtsov,
  JINR Rapid Commun. \textbf{2[76]},  5  (1996)
  [arXiv: hep-ph/9604363];
  Phys. Rev. Lett. \textbf{79},  1209  (1997);
  Theor. Math. Phys. \textbf{150},  132  (2007)
  [arXiv: hep-ph/0611229].

\bibitem{MS96}
 K.~A. Milton and I.~L. Solovtsov,
  Phys. Rev. \textbf{D55},  5295  (1997).

\bibitem{SS98}
 I.~L. Solovtsov and D.~V. Shirkov,
  Phys. Lett. \textbf{B442},  344  (1998).

\bibitem{BMS05}
 A.~P. Bakulev, S.~V. Mikhailov, and N.~G. Stefanis,
  Phys. Rev. \textbf{D72},  074014  (2005);
  Erratum:  Phys. Rev. \textbf{D72},  119908(E)  (2005).

\bibitem{BKS05}
 A.~P. Bakulev, A.~I. Karanikas, and N.~G. Stefanis,
  Phys. Rev. \textbf{D72},  074015  (2005).

\bibitem{BMS06}
 A.~P. Bakulev, S.~V. Mikhailov, and N.~G. Stefanis,
  Phys. Rev. \textbf{D75}, 056005  (2007).
  Erratum:  Phys. Rev. \textbf{D77},  079901(E)  (2008).

\bibitem{AB08}
 A.~P. Bakulev,
  arXiv:0805.0829 [hep-ph]
  (to be published in Phys. Part. Nucl.).

\bibitem{Rad82}
 A.~V. Radyushkin,
  JINR Rapid Commun. \textbf{78},  96  (1996).
   [JINR Preprint, E2-82-159, 26 Febr. 1982;
    arXiv: hep-ph/9907228].

\bibitem{KP82}
 N.~V. Krasnikov and A.~A. Pivovarov, 
  Phys. Lett. \textbf{B116},  168  (1982).

\bibitem{KS01}
 A.~I. Karanikas and N.~G. Stefanis,
  Phys. Lett. \textbf{B504},  225  (2001).
  Erratum:  Phys. Lett. \textbf{B636}, 330 (2006).

\bibitem{Mag99}
 B.~A. Magradze,
  Int. J. Mod. Phys. \textbf{A15},  2715  (2000).

\bibitem{CGHJK96}
 R. Corless {\it et~al.},
  Adv. Comput. Math. \textbf{5},  329  (1996).

\bibitem{Mag00}
 B.~A. Magradze,
  \uppercase{D}ubna preprint E2-2000-222, 2000
  [arXiv: hep-ph/0010070];
  \uppercase{P}reprint RMI-2003-55, 2003
  [arXiv: hep-ph/0305020].

\bibitem{FGOC81}
 R.~D. Field, R. Gupta, S. Otto, and L. Chang,
  Nucl. Phys. \textbf{B186},  429  (1981).

\bibitem{DR81}
 F.~M. Dittes and A.~V. Radyushkin,
  Sov. J. Nucl. Phys. \textbf{34},  293  (1981).

\bibitem{BT87}
 E. Braaten and S.-M. Tse,
  Phys. Rev. \textbf{D35},  2255  (1987).

\bibitem{MNP99a}
 B. Meli\'{c}, B. Ni\v{z}i\'{c}, and K. Passek,
  Phys. Rev. \textbf{D60},  074004  (1999).

\bibitem{Rad77}
 A.~V. Radyushkin,
  \uppercase{D}ubna preprint P2-10717, 1977
  [arXiv: hep-ph/0410276].

\bibitem{ER80}
 A.~V. Efremov and A.~V. Radyushkin,
  Phys. Lett. \textbf{B94},  245  (1980).

\bibitem{SSK99}
 N.~G. Stefanis, W. Schroers, and H.-C. Kim,
  Phys. Lett. \textbf{B449},  299  (1999);
  Eur. Phys. J. \textbf{C18},  137  (2000).

\bibitem{BPSS04}
 A.~P. Bakulev, K. Passek-Kumeri\v{c}ki, W. Schroers, and N.~G. Stefanis,
  Phys. Rev. \textbf{D70},  033014  (2004).
  Erratum: Phys. Rev. \textbf{D70}, 079906(E) (2004).

\bibitem{BLM83}
 S.~J. Brodsky, G.~P. Lepage, and P.~B. Mackenzie,
  Phys. Rev. \textbf{D28}, 228 (1983). 

\bibitem{BMS01}
 A.~P. Bakulev, S.~V. Mikhailov, and N.~G. Stefanis,
  Phys. Lett. \textbf{B508},  279  (2001).
  Erratum: Phys. Lett. \textbf{B590}, 309 (2004).

\bibitem{BP06}
 A.~P. Bakulev and A.~V. Pimikov,
  Acta Phys. Polon. \textbf{B37},  3627 (2006);
  PEPAN Lett. \textbf{4},  637  (2007)
  [arXiv: hep-ph/0608288].

\bibitem{AB06parus}
 A.~P. Bakulev,
   in {\em New Trends in High-Energy Physics,
   Proceedings of the Conference, Yalta (Crimea), 16--23 Sept., 2006},
   edited by P.~N. Bogolyubov {\it et~al.}
   (BITP NASU (Kiev), JINR (Dubna), Kiev, 2006), pp.\ 203--212
  [arXiv: hep-ph/0611139].

\bibitem{BCK05}
 P.~A. Baikov, K.~G. Chetyrkin, and J.~H. K{\"u}hn,
  Phys. Rev. Lett. \textbf{96},  012003  (2006).

\bibitem{BKM01}
 D.~J. Broadhurst, A.~L. Kataev, and C.~J. Maxwell,
  Nucl. Phys. \textbf{B592},  247  (2001).

\bibitem{KK08}
 A.~L. Kataev and V.~T. Kim,
  in {\em Proceedings of International Seminar on Contemporary Problems 
  of Elementary Particle Physics, Dedicated to the Memory of I.~L.~Sovtsov,
  Dubna, January 17--18, 2008.},
  edited by A.~P. Bakulev {\it et~al.} (JINR, Dubna, 2008), pp.\ 167--182
  [arXiv:0804.3992~(hep-ph)].

\bibitem{LEP07}
 J. Alcaraz {\it et~al.}, arXiv:0712.0929~(hep-ex).

\end{thebibliography}

\newcommand{\noopsort}[1]{} \newcommand{\printfirst}[2]{#1}
  \newcommand{\singleletter}[1]{#1} \newcommand{\switchargs}[2]{#2#1}

\end{document}